\documentstyle[12pt]{article}

\def\b{\begin{eqnarray}}
\def\e{\end{eqnarray}}
\def\Im{\mbox{Im}}

\def\PsfigVersion{1.10}
\def\setDriver{\DvipsDriver} 
\ifx\undefined\psfig\else\endinput\fi
\let\LaTeXAtSign=\@
\let\@=\relax
\edef\psfigRestoreAt{\catcode`\@=\number\catcode`@\relax}
\catcode`\@=11\relax
\newwrite\@unused
\def\ps@typeout#1{{\let\protect\string\immediate\write\@unused{#1}}}

\def\DvipsDriver{
	\ps@typeout{psfig/tex \PsfigVersion -dvips}
\def\PsfigSpecials{\DvipsSpecials} 	\def\ps@dir{/}
\def\ps@predir{} }
\def\OzTeXDriver{
	\ps@typeout{psfig/tex \PsfigVersion -oztex}
	\def\PsfigSpecials{\OzTeXSpecials}
	\def\ps@dir{:}
	\def\ps@predir{:}
	\catcode`\^^J=5
}

\def\figurepath{./:}

\def\DoPaths#1{\expandafter\EachPath#1\stoplist}
\def\leer{}
\def\EachPath#1:#2\stoplist{
  \ExistsFile{#1}{\SearchedFile}
  \ifx#2\leer
  \else
    \expandafter\EachPath#2\stoplist
  \fi}
\def\ps@dir{/}
\def\ExistsFile#1#2{%
   \openin1=\ps@predir#1\ps@dir#2
   \ifeof1
       \closein1
   \else
       \closein1
        \ifx\ps@founddir\leer
           \edef\ps@founddir{#1}
        \fi
   \fi}
\def\get@dir#1{%
  \def\ps@founddir{}
  \def\SearchedFile{#1}
  \DoPaths\figurepath
}

\def\@nnil{\@nil}
\def\@empty{}
\def\@psdonoop#1\@@#2#3{}
\def\@psdo#1:=#2\do#3{\edef\@psdotmp{#2}\ifx\@psdotmp\@empty \else
    \expandafter\@psdoloop#2,\@nil,\@nil\@@#1{#3}\fi}
\def\@psdoloop#1,#2,#3\@@#4#5{\def#4{#1}\ifx #4\@nnil \else
       #5\def#4{#2}\ifx #4\@nnil \else#5\@ipsdoloop #3\@@#4{#5}\fi\fi}
\def\@ipsdoloop#1,#2\@@#3#4{\def#3{#1}\ifx #3\@nnil 
       \let\@nextwhile=\@psdonoop \else
      #4\relax\let\@nextwhile=\@ipsdoloop\fi\@nextwhile#2\@@#3{#4}}
\def\@tpsdo#1:=#2\do#3{\xdef\@psdotmp{#2}\ifx\@psdotmp\@empty \else
    \@tpsdoloop#2\@nil\@nil\@@#1{#3}\fi}
\def\@tpsdoloop#1#2\@@#3#4{\def#3{#1}\ifx #3\@nnil 
       \let\@nextwhile=\@psdonoop \else
      #4\relax\let\@nextwhile=\@tpsdoloop\fi\@nextwhile#2\@@#3{#4}}
\ifx\undefined\fbox
\newdimen\fboxrule
\newdimen\fboxsep
\newdimen\ps@tempdima
\newbox\ps@tempboxa
\long\def\fbox#1{\leavevmode\setbox\ps@tempboxa\hbox{#1}\ps@tempdima\fboxrule
    \advance\ps@tempdima \fboxsep \advance\ps@tempdima \dp\ps@tempboxa
   \hbox{\lower \ps@tempdima\hbox
  {\vbox{\hrule height \fboxrule
          \hbox{\vrule width \fboxrule \hskip\fboxsep
          \vbox{\vskip\fboxsep \box\ps@tempboxa\vskip\fboxsep}\hskip 
                 \fboxsep\vrule width \fboxrule}
                 \hrule height \fboxrule}}}}
\fi
\newread\ps@stream
\newif\ifnot@eof       
\newif\if@noisy        
\newif\if@atend        
\newif\if@psfile       
{\catcode`\%=12\global\gdef\epsf@start{
\def\epsf@PS{PS}
\def\epsf@getbb#1{%
\openin\ps@stream=\ps@predir#1
\ifeof\ps@stream\ps@typeout{Error, File #1 not found}\else
   {\not@eoftrue \chardef\other=12
    \def\do##1{\catcode`##1=\other}\dospecials \catcode`\ =10
    \loop
       \if@psfile
	  \read\ps@stream to \epsf@fileline
       \else{
	  \obeyspaces
          \read\ps@stream to \epsf@tmp\global\let\epsf@fileline\epsf@tmp}
       \fi
       \ifeof\ps@stream\not@eoffalse\else
       \if@psfile\else
       \expandafter\epsf@test\epsf@fileline:. \\%
       \fi
          \expandafter\epsf@aux\epsf@fileline:. \\%
       \fi
   \ifnot@eof\repeat
   }\closein\ps@stream\fi}%
\long\def\epsf@test#1#2#3:#4\\{\def\epsf@testit{#1#2}
			\ifx\epsf@testit\epsf@start\else
\ps@typeout{Warning! File does not start with `\epsf@start'.  It may not be a PostScript file.}
			\fi
			\@psfiletrue} 
{\catcode`\%=12\global\let\epsf@percent=
\long\def\epsf@aux#1#2:#3\\{\ifx#1\epsf@percent
   \def\epsf@testit{#2}\ifx\epsf@testit\epsf@bblit
	\@atendfalse
        \epsf@atend #3 . \\%
	\if@atend	
	   \if@verbose{
		\ps@typeout{psfig: found `(atend)'; continuing search}
	   }\fi
        \else
        \epsf@grab #3 . . . \\%
        \not@eoffalse
        \global\no@bbfalse
        \fi
   \fi\fi}%
\def\epsf@grab #1 #2 #3 #4 #5\\{%
   \global\def\epsf@llx{#1}\ifx\epsf@llx\empty
      \epsf@grab #2 #3 #4 #5 .\\\else
   \global\def\epsf@lly{#2}%
   \global\def\epsf@urx{#3}\global\def\epsf@ury{#4}\fi}%
\def\epsf@atendlit{(atend)} 
\def\epsf@atend #1 #2 #3\\{%
   \def\epsf@tmp{#1}\ifx\epsf@tmp\empty
      \epsf@atend #2 #3 .\\\else
   \ifx\epsf@tmp\epsf@atendlit\@atendtrue\fi\fi}

\chardef\psletter = 11 
\chardef\other = 12

\newif \ifdebug 
\newif\ifc@mpute 
\c@mputetrue 

\let\then = \relax
\def\r@dian{pt }
\let\r@dians = \r@dian
\let\dimensionless@nit = \r@dian
\let\dimensionless@nits = \dimensionless@nit
\def\internal@nit{sp }
\let\internal@nits = \internal@nit
\newif\ifstillc@nverging
\def \Mess@ge #1{\ifdebug \then \message {#1} \fi}

{ 
	\catcode `\@ = \psletter
	\gdef \nodimen {\expandafter \n@dimen \the \dimen}
	\gdef \term #1 #2 #3%
	       {\edef \t@ {\the #1}
		\edef \t@@ {\expandafter \n@dimen \the #2\r@dian}%
		\t@rm {\t@} {\t@@} {#3}%
	       }
	\gdef \t@rm #1 #2 #3%
	       {{%
		\count 0 = 0
		\dimen 0 = 1 \dimensionless@nit
		\dimen 2 = #2\relax
		\Mess@ge {Calculating term #1 of \nodimen 2}%
		\loop
		\ifnum	\count 0 < #1
		\then	\advance \count 0 by 1
			\Mess@ge {Iteration \the \count 0 \space}%
			\Multiply \dimen 0 by {\dimen 2}%
			\Mess@ge {After multiplication, term = \nodimen 0}%
			\Divide \dimen 0 by {\count 0}%
			\Mess@ge {After division, term = \nodimen 0}%
		\repeat
		\Mess@ge {Final value for term #1 of 
				\nodimen 2 \space is \nodimen 0}%
		\xdef \Term {#3 = \nodimen 0 \r@dians}%
		\aftergroup \Term
	       }}
	\catcode `\p = \other
	\catcode `\t = \other
	\gdef \n@dimen #1pt{#1} 
}

\def \Divide #1by #2{\divide #1 by #2} 

\def \Multiply #1by #2
       {{
	\count 0 = #1\relax
	\count 2 = #2\relax
	\count 4 = 65536
	\Mess@ge {Before scaling, count 0 = \the \count 0 \space and
			count 2 = \the \count 2}%
	\ifnum	\count 0 > 32767 
	\then	\divide \count 0 by 4
		\divide \count 4 by 4
	\else	\ifnum	\count 0 < -32767
		\then	\divide \count 0 by 4
			\divide \count 4 by 4
		\else
		\fi
	\fi
	\ifnum	\count 2 > 32767 
	\then	\divide \count 2 by 4
		\divide \count 4 by 4
	\else	\ifnum	\count 2 < -32767
		\then	\divide \count 2 by 4
			\divide \count 4 by 4
		\else
		\fi
	\fi
	\multiply \count 0 by \count 2
	\divide \count 0 by \count 4
	\xdef \product {#1 = \the \count 0 \internal@nits}%
	\aftergroup \product
       }}

\def\r@duce{\ifdim\dimen0 > 90\r@dian \then   
		\multiply\dimen0 by -1
		\advance\dimen0 by 180\r@dian
		\r@duce
	    \else \ifdim\dimen0 < -90\r@dian \then  
		\advance\dimen0 by 360\r@dian
		\r@duce
		\fi
	    \fi}

\def\Sine#1%
       {{%
	\dimen 0 = #1 \r@dian
	\r@duce
	\ifdim\dimen0 = -90\r@dian \then
	   \dimen4 = -1\r@dian
	   \c@mputefalse
	\fi
	\ifdim\dimen0 = 90\r@dian \then
	   \dimen4 = 1\r@dian
	   \c@mputefalse
	\fi
	\ifdim\dimen0 = 0\r@dian \then
	   \dimen4 = 0\r@dian
	   \c@mputefalse
	\fi
	\ifc@mpute \then
		\divide\dimen0 by 180
		\dimen0=3.141592654\dimen0
		\dimen 2 = 3.1415926535897963\r@dian 
		\divide\dimen 2 by 2 
		\Mess@ge {Sin: calculating Sin of \nodimen 0}%
		\count 0 = 1 
		\dimen 2 = 1 \r@dian 
		\dimen 4 = 0 \r@dian 
		\loop
			\ifnum	\dimen 2 = 0 
			\then	\stillc@nvergingfalse 
			\else	\stillc@nvergingtrue
			\fi
			\ifstillc@nverging 
			\then	\term {\count 0} {\dimen 0} {\dimen 2}%
				\advance \count 0 by 2
				\count 2 = \count 0
				\divide \count 2 by 2
				\ifodd	\count 2 
				\then	\advance \dimen 4 by \dimen 2
				\else	\advance \dimen 4 by -\dimen 2
				\fi
		\repeat
	\fi		
			\xdef \sine {\nodimen 4}%
       }}

\def\Cosine#1{\ifx\sine\UnDefined\edef\Savesine{\relax}\else
		             \edef\Savesine{\sine}\fi
	{\dimen0=#1\r@dian\advance\dimen0 by 90\r@dian
	 \Sine{\nodimen 0}
	 \xdef\cosine{\sine}
	 \xdef\sine{\Savesine}}}	      

\def\psdraft{
	\def\@psdraft{0}
}
\def\psfull{
	\def\@psdraft{100}
}

\psfull

\newif\if@scalefirst
\def\psscalefirst{\@scalefirsttrue}
\def\psrotatefirst{\@scalefirstfalse}
\psrotatefirst

\newif\if@draftbox
\def\psnodraftbox{
	\@draftboxfalse
}
\def\psdraftbox{
	\@draftboxtrue
}
\@draftboxtrue

\newif\if@prologfile
\newif\if@postlogfile
\def\pssilent{
	\@noisyfalse
}
\def\psnoisy{
	\@noisytrue
}
\psnoisy
\newif\if@bbllx
\newif\if@bblly
\newif\if@bburx
\newif\if@bbury
\newif\if@height
\newif\if@width
\newif\if@rheight
\newif\if@rwidth
\newif\if@angle
\newif\if@clip
\newif\if@verbose
\def\@p@@sclip#1{\@cliptrue}
\newif\if@decmpr
\def\@p@@sfigure#1{\def\@p@sfile{null}\def\@p@sbbfile{null}\@decmprfalse
   \openin1=\ps@predir#1
   \ifeof1
	\closein1
	\get@dir{#1}
	\ifx\ps@founddir\leer
		\openin1=\ps@predir#1.bb
		\ifeof1
			\closein1
			\get@dir{#1.bb}
			\ifx\ps@founddir\leer
				\ps@typeout{Can't find #1 in \figurepath}
			\else
				\@decmprtrue
				\def\@p@sfile{\ps@founddir\ps@dir#1}
				\def\@p@sbbfile{\ps@founddir\ps@dir#1.bb}
			\fi
		\else
			\closein1
			\@decmprtrue
			\def\@p@sfile{#1}
			\def\@p@sbbfile{#1.bb}
		\fi
	\else
		\def\@p@sfile{\ps@founddir\ps@dir#1}
		\def\@p@sbbfile{\ps@founddir\ps@dir#1}
	\fi
   \else
	\closein1
	\def\@p@sfile{#1}
	\def\@p@sbbfile{#1}
   \fi
}
\def\@p@@sfile#1{\@p@@sfigure{#1}}
\def\@p@@sbbllx#1{
		\@bbllxtrue
		\dimen100=#1
		\edef\@p@sbbllx{\number\dimen100}
}
\def\@p@@sbblly#1{
		\@bbllytrue
		\dimen100=#1
		\edef\@p@sbblly{\number\dimen100}
}
\def\@p@@sbburx#1{
		\@bburxtrue
		\dimen100=#1
		\edef\@p@sbburx{\number\dimen100}
}
\def\@p@@sbbury#1{
		\@bburytrue
		\dimen100=#1
		\edef\@p@sbbury{\number\dimen100}
}
\def\@p@@sheight#1{
		\@heighttrue
		\dimen100=#1
   		\edef\@p@sheight{\number\dimen100}
}
\def\@p@@swidth#1{
		\@widthtrue
		\dimen100=#1
		\edef\@p@swidth{\number\dimen100}
}
\def\@p@@srheight#1{
		\@rheighttrue
		\dimen100=#1
		\edef\@p@srheight{\number\dimen100}
}
\def\@p@@srwidth#1{
		\@rwidthtrue
		\dimen100=#1
		\edef\@p@srwidth{\number\dimen100}
}
\def\@p@@sangle#1{
		\@angletrue
		\edef\@p@sangle{#1} 
}
\def\@p@@ssilent#1{ 
		\@verbosefalse
}
\def\@p@@sprolog#1{\@prologfiletrue\def\@prologfileval{#1}}
\def\@p@@spostlog#1{\@postlogfiletrue\def\@postlogfileval{#1}}
\def\@cs@name#1{\csname #1\endcsname}
\def\@setparms#1=#2,{\@cs@name{@p@@s#1}{#2}}
\def\ps@init@parms{
		\@bbllxfalse \@bbllyfalse
		\@bburxfalse \@bburyfalse
		\@heightfalse \@widthfalse
		\@rheightfalse \@rwidthfalse
		\def\@p@sbbllx{}\def\@p@sbblly{}
		\def\@p@sbburx{}\def\@p@sbbury{}
		\def\@p@sheight{}\def\@p@swidth{}
		\def\@p@srheight{}\def\@p@srwidth{}
		\def\@p@sangle{0}
		\def\@p@sfile{} \def\@p@sbbfile{}
		\def\@p@scost{10}
		\def\@sc{}
		\@prologfilefalse
		\@postlogfilefalse
		\@clipfalse
		\if@noisy
			\@verbosetrue
		\else
			\@verbosefalse
		\fi
}
\def\parse@ps@parms#1{
	 	\@psdo\@psfiga:=#1\do
		   {\expandafter\@setparms\@psfiga,}}
\newif\ifno@bb
\def\bb@missing{
	\if@verbose{
		\ps@typeout{psfig: searching \@p@sbbfile \space  for bounding box}
	}\fi
	\no@bbtrue
	\epsf@getbb{\@p@sbbfile}
        \ifno@bb \else \bb@cull\epsf@llx\epsf@lly\epsf@urx\epsf@ury\fi
}	
\def\bb@cull#1#2#3#4{
	\dimen100=#1 bp\edef\@p@sbbllx{\number\dimen100}
	\dimen100=#2 bp\edef\@p@sbblly{\number\dimen100}
	\dimen100=#3 bp\edef\@p@sbburx{\number\dimen100}
	\dimen100=#4 bp\edef\@p@sbbury{\number\dimen100}
	\no@bbfalse
}
\newdimen\p@intvaluex
\newdimen\p@intvaluey
\def\rotate@#1#2{{\dimen0=#1 sp\dimen1=#2 sp
		  \global\p@intvaluex=\cosine\dimen0
		  \dimen3=\sine\dimen1
		  \global\advance\p@intvaluex by -\dimen3
		  \global\p@intvaluey=\sine\dimen0
		  \dimen3=\cosine\dimen1
		  \global\advance\p@intvaluey by \dimen3
		  }}
\def\compute@bb{
		\no@bbfalse
		\if@bbllx \else \no@bbtrue \fi
		\if@bblly \else \no@bbtrue \fi
		\if@bburx \else \no@bbtrue \fi
		\if@bbury \else \no@bbtrue \fi
		\ifno@bb \bb@missing \fi
		\ifno@bb \ps@typeout{FATAL ERROR: no bb supplied or found}
			\no-bb-error
		\fi
		\count203=\@p@sbburx
		\count204=\@p@sbbury
		\advance\count203 by -\@p@sbbllx
		\advance\count204 by -\@p@sbblly
		\edef\ps@bbw{\number\count203}
		\edef\ps@bbh{\number\count204}
		\if@angle 
			\Sine{\@p@sangle}\Cosine{\@p@sangle}
	        	{\dimen100=\maxdimen\xdef\r@p@sbbllx{\number\dimen100}
					    \xdef\r@p@sbblly{\number\dimen100}
			                    \xdef\r@p@sbburx{-\number\dimen100}
					    \xdef\r@p@sbbury{-\number\dimen100}}
                        \def\minmaxtest{
			   \ifnum\number\p@intvaluex<\r@p@sbbllx
			      \xdef\r@p@sbbllx{\number\p@intvaluex}\fi
			   \ifnum\number\p@intvaluex>\r@p@sbburx
			      \xdef\r@p@sbburx{\number\p@intvaluex}\fi
			   \ifnum\number\p@intvaluey<\r@p@sbblly
			      \xdef\r@p@sbblly{\number\p@intvaluey}\fi
			   \ifnum\number\p@intvaluey>\r@p@sbbury
			      \xdef\r@p@sbbury{\number\p@intvaluey}\fi
			   }
			\rotate@{\@p@sbbllx}{\@p@sbblly}
			\minmaxtest
			\rotate@{\@p@sbbllx}{\@p@sbbury}
			\minmaxtest
			\rotate@{\@p@sbburx}{\@p@sbblly}
			\minmaxtest
			\rotate@{\@p@sbburx}{\@p@sbbury}
			\minmaxtest
			\edef\@p@sbbllx{\r@p@sbbllx}\edef\@p@sbblly{\r@p@sbblly}
			\edef\@p@sbburx{\r@p@sbburx}\edef\@p@sbbury{\r@p@sbbury}
		\fi
		\count203=\@p@sbburx
		\count204=\@p@sbbury
		\advance\count203 by -\@p@sbbllx
		\advance\count204 by -\@p@sbblly
		\edef\@bbw{\number\count203}
		\edef\@bbh{\number\count204}
}
\def\in@hundreds#1#2#3{\count240=#2 \count241=#3
		     \count100=\count240	
		     \divide\count100 by \count241
		     \count101=\count100
		     \multiply\count101 by \count241
		     \advance\count240 by -\count101
		     \multiply\count240 by 10
		     \count101=\count240	
		     \divide\count101 by \count241
		     \count102=\count101
		     \multiply\count102 by \count241
		     \advance\count240 by -\count102
		     \multiply\count240 by 10
		     \count102=\count240	
		     \divide\count102 by \count241
		     \count200=#1\count205=0
		     \count201=\count200
			\multiply\count201 by \count100
		 	\advance\count205 by \count201
		     \count201=\count200
			\divide\count201 by 10
			\multiply\count201 by \count101
			\advance\count205 by \count201
		     \count201=\count200
			\divide\count201 by 100
			\multiply\count201 by \count102
			\advance\count205 by \count201
		     \edef\@result{\number\count205}
}
\def\compute@wfromh{
		\in@hundreds{\@p@sheight}{\@bbw}{\@bbh}
		\edef\@p@swidth{\@result}
}
\def\compute@hfromw{
	        \in@hundreds{\@p@swidth}{\@bbh}{\@bbw}
		\edef\@p@sheight{\@result}
}
\def\compute@handw{
		\if@height 
			\if@width
			\else
				\compute@wfromh
			\fi
		\else 
			\if@width
				\compute@hfromw
			\else
				\edef\@p@sheight{\@bbh}
				\edef\@p@swidth{\@bbw}
			\fi
		\fi
}
\def\compute@resv{
		\if@rheight \else \edef\@p@srheight{\@p@sheight} \fi
		\if@rwidth \else \edef\@p@srwidth{\@p@swidth} \fi
}
\def\compute@sizes{
	\compute@bb
	\if@scalefirst\if@angle
	\if@width
	   \in@hundreds{\@p@swidth}{\@bbw}{\ps@bbw}
	   \edef\@p@swidth{\@result}
	\fi
	\if@height
	   \in@hundreds{\@p@sheight}{\@bbh}{\ps@bbh}
	   \edef\@p@sheight{\@result}
	\fi
	\fi\fi
	\compute@handw
	\compute@resv}
\def\OzTeXSpecials{
	\special{empty.ps /@isp {true} def}
	\special{empty.ps \@p@swidth \space \@p@sheight \space
			\@p@sbbllx \space \@p@sbblly \space
			\@p@sbburx \space \@p@sbbury \space
			startTexFig \space }
	\if@clip{
		\if@verbose{
			\ps@typeout{(clip)}
		}\fi
		\special{empty.ps doclip \space }
	}\fi
	\if@angle{
		\if@verbose{
			\ps@typeout{(rotate)}
		}\fi
		\special {empty.ps \@p@sangle \space rotate \space} 
	}\fi
	\if@prologfile
	    \special{\@prologfileval \space } \fi
	\if@decmpr{
		\if@verbose{
			\ps@typeout{psfig: Compression not available
			in OzTeX version \space }
		}\fi
	}\else{
		\if@verbose{
			\ps@typeout{psfig: including \@p@sfile \space }
		}\fi
		\special{epsf=\@p@sfile \space }
	}\fi
	\if@postlogfile
	    \special{\@postlogfileval \space } \fi
	\special{empty.ps /@isp {false} def}
}
\def\DvipsSpecials{
	\special{ps::[begin] 	\@p@swidth \space \@p@sheight \space
			\@p@sbbllx \space \@p@sbblly \space
			\@p@sbburx \space \@p@sbbury \space
			startTexFig \space }
	\if@clip{
		\if@verbose{
			\ps@typeout{(clip)}
		}\fi
		\special{ps:: doclip \space }
	}\fi
	\if@angle
		\if@verbose{
			\ps@typeout{(clip)}
		}\fi
		\special {ps:: \@p@sangle \space rotate \space} 
	\fi
	\if@prologfile
	    \special{ps: plotfile \@prologfileval \space } \fi
	\if@decmpr{
		\if@verbose{
			\ps@typeout{psfig: including \@p@sfile.Z \space }
		}\fi
		\special{ps: plotfile "`zcat \@p@sfile.Z" \space }
	}\else{
		\if@verbose{
			\ps@typeout{psfig: including \@p@sfile \space }
		}\fi
		\special{ps: plotfile \@p@sfile \space }
	}\fi
	\if@postlogfile
	    \special{ps: plotfile \@postlogfileval \space } \fi
	\special{ps::[end] endTexFig \space }
}
\def\psfig#1{\vbox {
	\ps@init@parms
	\parse@ps@parms{#1}
	\compute@sizes
	\ifnum\@p@scost<\@psdraft{
		\PsfigSpecials 
		\vbox to \@p@srheight sp{
			\hbox to \@p@srwidth sp{
				\hss
			}
		\vss
		}
	}\else{
		\if@draftbox{		
			\hbox{\fbox{\vbox to \@p@srheight sp{
			\vss
			\hbox to \@p@srwidth sp{ \hss 
			 \hss }
			\vss
			}}}
		}\else{
			\vbox to \@p@srheight sp{
			\vss
			\hbox to \@p@srwidth sp{\hss}
			\vss
			}
		}\fi

	}\fi
}}
\psfigRestoreAt
\setDriver
\let\@=\LaTeXAtSign

\begin{document}

\pagestyle{myheadings}

\title{{\normalsize quant-ph/9604002 
\hfill Z\"urich University Preprint:
ZU--TH--10/96}\\[0.3cm]
Multiphoton Ionization as Time--Dependent Tunneling}

\author{\\Klaus Ergenzinger\thanks{e-mail: erg@physik.unizh.ch}\\[0.2cm]
Institut f\"ur Theoretische Physik\\ Universit\"at
Z\"urich\\ Winterthurerstr. 190, CH-8057 Z\"urich, Switzerland\\}

\date{April 3, 1996} 

\maketitle

\begin{abstract}

A new semiclassical approach to ionization by an oscillating
field is presented.
For a $\delta$--function atom, an asymptotic analysis is performed with
respect to a
quantity $h$, defined as the ratio of photon
energy to ponderomotive energy. This $h$ appears formally equivalent to
Planck's
constant in a suitably transformed Schr\"odinger equation and allows
semiclassical
methods to be applicable. Systematically, a picture of
tunneling wave packets in complex time
is developped, which by interference account for the typical
ponderomotive features of
ionization curves. These analytical results are then compared
to numerical simulations \cite{ksdiss}  and are shown to be in good
agreement.

\end{abstract}

\clearpage
\section{Introduction}
In recent times, there has been a lot of effort dedicated to a better
understanding of ionization by strong laser fields (reviews e.g. in
\cite{mainfray,knightreview}), especially since the discovery of
nonperturbative effects like above--threshold ionization ATI (for a
review see \cite{eberlyrev}), the sensitivity of ionization rates, and
stabilization in superintense fields \cite{gavrila,special,ks2}.
But there succeeded no analytical solution, not even to the simplest
model, i.e.
an electron bound by an attractive $\delta$--potential in the presence
of an oscillating electric field.
One of the main reasons is that
there exist two separate regions. Inside the atomic core, the
binding potential is predominant, and outside, it is vice versa
the electric field that dominates
(this is also the main point that makes perturbation theory work
so poorly).
For both regions, the distinct propagators are known exactly, but they
cannot
be combined to solve the ionization problem exactly.

An appropriate way of addressing this problem nonperturbatively is the
semiclassical method, which we will use to construct the
semiclassical propagator for ionization of such a $\delta$--function
atom.
The choice of this model has three advantages: first, there is a clear
distinction
between inside and outside the atom, so that there exists no
intermediate region.
Second, there is only one bound state for the $\delta$--potential, so
there
arise no difficulties with intermediate resonances and induced
resonances by AC--Stark shift as it happens in real atoms. In this
point, our model
resembles photodetachment of a negative ion \cite{negions}, especially
$\mbox{H}^-$. Third, the exact problem can be reduced to a Volterra
type integral equation in time,
for which accurate numerical solutions can be computed \cite{ksdiss} and
allow precise tests.

The above described problem (as well as its more general settings in
three dimensions with more realistic binding potentials) has been
treated in the literature in several ways.
The so--called Keldysh--Faisal--Reiss (KFR) approach  \cite{keldysh,
faisal,
zreiss} consists in expressing the exact propagator $\hat{U}$ in terms
of the known Volkov propagator $\hat{U}^V$ \cite{volkov}
(for a free electron in the electromagnetic field) and of $V_{\delta}$, the
atomic binding potential:
$\hat{U}=\hat{U}^V+\hat{U}^V\,V_{\delta}\,\hat{U}$. The unknown
$\hat{U}$ on the right hand side is approximated by $\hat{U}^V$ (equivalent
to Born approximation), and matrix--elements for ionization  are calculated
between the ground state and so--called Volkov--states \cite{volkov} in
the continuum. This approach has been extended
and refined by many authors, and we will compare our results with two
such typical extensions \cite{perelomovbasic,susskind}.

Another approach, the so--called two--step model
\cite{scully,kulander,corkum}, clearly distinguishes between ionization
first and classical propagation in the laser field afterwards. This
proved to be very useful especially in calculating high--harmonic
generation \cite{lewenstein}. This separation into two steps will be
used in the following, but now justified in a fully semiclassical context.

In addition, there exist several other approaches; a very common method
is using Floquet theory \cite{shirley,chu,floquetposha}, which
explicitly incorporates the periodicity of the time--dependent
Hamiltonian.

Our issue here is not to obtain better results for a simple model, but
to gain better physical insight into the mechanisms of ionization processes
using semiclassical methods.

This paper is organized as follows:
After basic definitions, we demonstrate characteristic elements of
typical
ionization curves which we want to understand semiclassically.
Using a ``sum over classical paths'' technique, the total semiclassical
propagator
is constructed successively by identifying the paths which are relevant
for ionization. First for the part remaining bound, and then for the
free wave packets stemming from time--dependent tunneling in complex
time.
From this semiclassical propagator, the total ionization rate is
derived and compared to numerical simulations as well as to other
theories in the literature.

\clearpage

\section{Basics}

\subsection{Definition of the model}
\label{model}
We want to study the ionization of a one--dimensional
$\delta$--function atom $V_\delta  =-\alpha \delta(x)$, which possesses
exactly one bound state with energy ${E_0=-\alpha ^2/2}$ (for the
three--dimensional analog cf. \cite{beckerdreid}).
This atom is exposed to an oscillating electric field
$V_0=-\mu x\cos(\omega t)$ in the so--called dipole--approximation.
The parameter $\alpha $ stands for the strength of the binding
potential, $\mu$ for the amplitude of the applied electric field, and
$\omega$ for its angular frequency. The Schr\"odinger equation in
atomic units ($\hbar=m=1$)
is
\b i{\partial \over \partial t}\,\psi=\bigg(-{1 \over 2} {\partial ^2
\over \partial x^2}-\alpha
\delta (x)-\mu x \cos(\omega t)\bigg)\,\psi \label{schroedinger}\e

From these three parameters, there can be derived (due to the scaling
properties of (\ref{schroedinger})) two independent, meaningful, and
dimensionless quantities:

\begin{itemize}
\item[]first, there is $z$, the ratio of ponderomotive
energy $U_{pond}=\mu^2 / ( 4 \omega^2)$ to
photon energy $\omega$: ${z={\mu^2 /( 4 \omega^3)}}$.
$U_{pond}$ is the mean kinetic energy of a free electron in an
oscillating field.
\item[] and second the so--called Keldysh--factor \cite{keldysh}
${\gamma={\alpha \omega / \mu}}$, which is the ratio of the (adiabatic)
tunneling time to the period of the applied field. This $\gamma$
characterizes the
ionization process; $\gamma \ll 1$ corresponds to (adiabatic) tunneling,
and $\gamma \gg 1$ is better described in a pure multiphoton frame
\cite{delonekrainov,faisalbook}.
\end{itemize}

These are the two quantities by which the model will be described
below. A third, but no more
independent quantity is $n_{io}$, the ratio
of necessary ionization energy $E_0$ to photon
energy:
\b n_{io}={\alpha^2 / (2 \omega)}=2 \gamma^2 z \label{nio} \e

\subsection{Transformation of the Schr\"odinger equation}
\label{trafo}
Using the following coordinate and time transformation, we cast the
above
Schr\"odinger equation (\ref{schroedinger}) in a very suitable form:
\b x'={\omega^2\over \mu}x\ ,\     t'=\omega t 
\label{hdef} \e
Using the scaling relation $\delta(a\,x)=\delta(x)/|a|$, which is
identical to the
scaling behaviour of the Coulomb--potential, and omitting the primes,
we arrive at:
\b i h{\partial \over \partial t}\,\psi=\bigg(-{1 \over 2}h^2
{\partial ^2 \over \partial x^2}
-h \gamma \delta (x)- x \cos(t)\bigg)\,\psi \label{transformed} \e
Here ${h:=\omega^3/\mu^2={1 / (4\, z)}}$ is written very suggestively to
indicate that we
have obtained a parameter $h$ formally identical to Planck's constant
$\hbar$
in ordinary quantum mechanics using SI units. Of course we can give $h$
any
value we like. Restricting ourselves to strong fields where
$U_{pond}\gg \omega$, i.e.
$z\gg 1$, $h$ can get arbitrarily small. This will allow us to use the
normal semiclassical methods, exploiting the formal analogy between
the parameter $h$, introduced above, and Planck's constant $\hbar$.
Still remarkable is the appearance of the factor $h$ in front of the
binding $\delta$--potential, its implications in the semiclassical
limit will be derived below.


The (normalized) ground state wavefunction for the $\delta$--function
atom
without applied external field is
\b \psi_0(x)&=&\sqrt{{\gamma \over h}}\exp(-{\gamma\over h}|x|)
\label{psinull} \\
\hat{H}_0\,\psi_0&=&-{\gamma^2\over 2}\psi_0\\
\psi_0(x,t) &=&\psi_0(x)\exp(i{\gamma^2\over2h}t)\ , \nonumber \e

which leads to the stronger a localization the smaller $h$ is. In the
limiting case $h \to 0$,
$\psi_0(x)$ approaches (appropriately scaled) the spatial Dirac
$\delta$--function:
\b {1 \over 2} \sqrt{\gamma \over h}\,\psi_0(x)\to \delta(x)
\label{spatialdelta} \e

In the following, this approximation will be used only for calculation
of
scalar products, so that no mathematical ambiguities will arise.

\section{Prominent Semiclassical Features in Ionization Rates}
\begin{figure}[htb]
\psfig{file=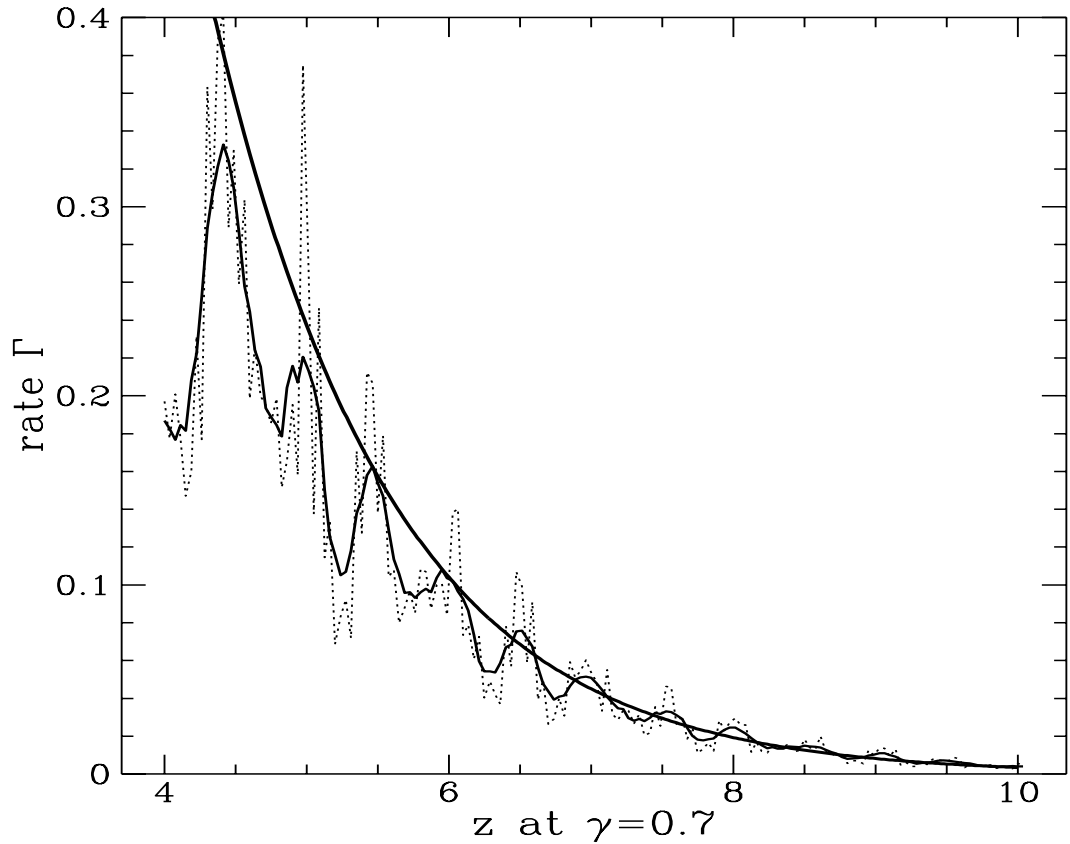,width=15cm,height= 12cm}
\caption[Numerical ionization rates, raw (dotted) and smoothed
(thicker) versus number of ponderomotive photons
$  z$, compared with WKB--background (thick, smooth decaying
curve).]
{Numerical ionization rates, raw (dotted) and smoothed
(thicker) versus number of
ponderomotive photons
$  z$, compared with WKB--background (thick, smooth decaying
curve). The Keldysh parameter $\gamma$ and not the depth of the binding
potential is kept fixed.}
\label{prominentsemi}
\end{figure}
The ionization rates from a numerical analysis \cite{ksdiss} of the
model show certain characteristic features. In figure \ref{prominentsemi}, we
see the raw results (dotted), these results smoothed (thicker)  and the
adiabatically averaged WKB--value (see eq. (\ref{dquer}), depicted thick,
monotonically decreasing). The ionization rate $\Gamma$ is shown
against
$z$, the ponderomotive energy (over photon energy). The WKB--rate accounts
well for
the background, but on the actual rate, there is a periodic modulation
and a lot
of fine--structure superposed. The (slow) modulation becomes obvious
after smoothing the raw data (using a Savitzky--Golay filter technique
\cite{recipes})
in order to eliminate the fine structure
and to work out this fundamental modulation.

The period of this
modulation can be understood by so--called channel closing arguments
\cite{eberlychannel}, but there is no argument for the amplitude of
modulation. The $k$--th channel means a multiphoton ionization by $k$
photons with energy balance (here in atomic units)
\b k\,\omega=n_{io}\,\omega + z\,\omega + E_{kin} \label{bilance}\e
$n_{io}\,\omega$ is the energy required for ionization, $z\,\omega$ is
the (ponderomotive) energy $U_{pond}$ contained in the electron
oscillation in the electric field, and $E_{kin}$ is the additional
kinetic energy the free electron gains in this ionization process.
This channel is energetically only allowed for $E_{kin}\geq 0$,
otherwise it is  forbidden.

The threshold for the $k$--th channel is defined using the condition
$E_{kin}=0$. In terms of $\gamma$ and $z$, equation (\ref{bilance}) at
threshold is
\b k=2\gamma^2 z_k + z_k \e
and the specific value $z_k$ at threshold becomes
\b z_k={k \over 1+2\gamma^2}\ , \label{zks} \e
yielding a period $\Delta z=1/(1+2\gamma^2)$ in $z$, i.e. the period of
approximately one half in the example of figure \ref{prominentsemi}.

An important point to note is that the numerical results give strong
evidence for regular behaviour at threshold, whereas the usual
prediction of appropriate theories (e.g. those of KFR type) is a
divergence at threshold.

Here it is important to note that the features described above are definitely
not restricted only to the $\delta$--function atom. Numerical
simulations for various model potentials in the literature exhibit similar
features, often with remarkable quantitative correspondence (see figure 8
in \cite{ksdiss}, comparing the $\delta$--function atom to a smoothed binding
potential $V(x)= -\exp(-|x|)/\sqrt{x^2+x_0^2}$ as used by Greenwood
and Eberly \cite{ebergreen}).

In the following, a semiclassical theory is derived
that accounts for the information contained in smoothed rates. The
WKB--background as well as the periodicity, amplitude, and phase of
modulation are contained in a single, divergence free theory, which
constructs the propagator
using the semiclassical sum over paths, cf. e.g.
\cite{gutzwillerbook}.

\section{Quasi--Energies}

\subsection{Calculation of WKB-coefficient}
The main effect of applying an external field to an atom is
that
the bound state becomes metastable and tunneling can occur.
In the static case, this tunneling rate $D$ can be approximated using
the
normal WKB--coefficient 
for the corresponding barrier.
For a linear potential barrier $V(x)=-\eta x$ and a given (negative)
energy
$E_0$,
the well--known expression for $D$ is
\b D(\eta)\approx\exp\bigg(-{2\over
h}\!\int\limits_0^{-E_0/\eta}|p(x)|dx\bigg) \e
Using the (imaginary) local momentum $p(x)=\sqrt{2(E_0+\eta x)}$ and
the ground state energy $E_0=-\gamma^2/2$, this evaluates to
\b D(\eta)\approx\exp\bigg(-{2\over 3}{\gamma^3\over \eta
\,h}\bigg) \e

Of course, this approach gives only the exponential part, but since the
preexponential part is well--known from the literature for this simple
case (e.g.
\cite{susskind,elberfeldkleber}), we can adopt it from there. This
factor is just twice the atomic frequency, in our units $2 E_0/h$
\b D(\eta)={\gamma^2\over h}\exp\bigg(-{2\over 3}{\gamma^3\over
\eta \,h}\bigg) \e

If we consider a time--dependent external field, the parameter $\eta$
becomes time--dependent too and represents the instantaneous strength
of the applied field at the moment the ionization takes place:
$\eta=|\cos(t)|$. If the tunneling process occurs on a much
shorter time scale than the period $2\pi$ of the oscillation, it is a
good idea to consider the ionization taking place adiabatically, i.e.
calculate the instantaneous
ionization rate $D(|\cos(t)|)$ and average it over a whole period. This
case corresponds to $\gamma \ll 1$, i.e. the Keldysh factor
characterizing the ionization is quite small.

In calculating the cycle--average
$\bar{D}$ over a cosine period,
one has to integrate and to normalize subsequently
\b \bar{D}={1\over 2 \pi}\int\limits_0^{2\pi}{\gamma^2\over
h}\exp\bigg(-{2\over 3}{\gamma^3\over |\cos(t)| \,h}\bigg)dt \e

Because we want to examine the asymptotic case $h \to 0$, we best
evaluate this
integral using the method of steepest descent 
(also called saddle--point integration, cf. appendix \ref{spi})
with respect to $h$ . The
derivation of the exponential argument with respect to time $t$ gives
\b {d \over dt}\bigg(-{2\over 3}{\gamma^3\over \cos(t)}\bigg)=-{2\over
3}\gamma^3{\sin(t) \over {\cos(t)}^2} \e
The relevant times are the zeros of this expression, i.e. all multiples
of $\pi$. These are the instants where the electric field strength
$|\cos(t)|$ is a maximum. For symmetry reasons, all these instants are
equivalent
and it is sufficient to evaluate the above integral around one such
instant using the method of steepest descent. The result is
\b \bar{D}&=&\sqrt{3 h \over \pi \gamma^3}\,{\gamma^2\over
h}\exp\bigg(-{2\over 3}{\gamma^3\over h}\bigg) \label{dquer}\\
	 &=&\sqrt{3 h \over \pi \gamma^3}\,D(\eta\!=\!1) \e
This shows that the average $\bar{D}$ is the instantaneous ionization
rate at maximum field strength, up to a preexponential factor.

Note
that the method of steepest descent becomes exact in the limit of
vanishing $h$. This allows a very interesting interpretation.
In this case, the ionization
effectively takes place only in the vicinity of the instants of maximal
field strength $\eta=1$.
This means that there exist ionization bursts, separated by half a
period, between which practically no other ionization occurs.
In the following, this property will be used to construct a scenario of
propagating  wave packets, emerging from these peak instants, freely
propagating afterwards, and interfering with one another and with the
part of the wave function remaining bound.

\subsection{AC--Stark effect}
The second effect of applying an external field is the so--called Stark
effect,
in the oscillating time-dependent case called AC--Stark effect .
We will treat this effect adiabatically as well (cf. \cite{pontshake}),
the well--known value
(e.g. \cite{susskind,elberfeldkleber,geltman}) for the instantaneous
energy--shift $E^{AC}$ is
$E^{AC}(\eta)= -{5}{h^2 \eta^2 /( 8 \gamma^4)}$.
Cycle--averaging results in
\b \bar{E}^{AC}={1\over 2}E^{AC}(\eta\!=\!1)=-{5h^2\over 16\gamma^4}
\label{eacbar} \e
This effect means an additional phase factor in the propagator
$\exp(-i\hat{H}t/h)$, whereas the tunneling rate, calculated by taking
absolute squares of the wave function, is not affected by $E^{AC}$.
Note that the influence of (\ref{eacbar}) will be quite small in the
following because of its proportionality to $h^2$.

In order to express tunneling and Stark shift together, it is useful to
write
the exponential decay of the bound state using an imaginary
contribution $E^{I}$  to the total energy $E^m$. Setting
\b E^{I}(\eta) &=& -i{\gamma^2 \over 2}\exp\bigg(-{2 \over
3}{\gamma^3\over h\,\eta}\bigg)=-{i\, h\over 2}\,D(\eta)\\
 E^{m}(\eta) &=& E_0 + E^{AC}(\eta)+E^{I}(\eta)\ ,
\e
the adiabatic development of the ground state can be described by the
propagator
\b \hat{U}^\delta (t_f) &=& \exp\bigg(-{i\over h}\int\limits_0^{t_f}
E^{m}(|\cos(t)|)\,dt\bigg) \ \hat{P}_0\ , \e
using the total quasi--energy $E^{m}$ and the projection operator $\hat{P}_0$,
projecting onto the ground state.
This adiabatic description is useful if we want to describe the
propagation of the wavefunction for arbitrary times. If we restrict
ourselves to considering only full cycles, we can use the appropriate
averages.
\b   \bar{E}^{I} &=& \sqrt{3 h\over \pi \gamma^3} E^{I}(\eta\!=\!1)
\\
 \bar{E}^{m} &=& E_0 + \bar{E}^{AC}+\bar{E}^{I}  \label{barem}\\
  \hat{\bar{U}}^\delta (t_f) &=&  \exp\bigg(-{i\over h}\bar{E}^{m}\,
  t_f\bigg)\, \hat{P}_0 \label{udeltabar}\e
Applying this propagator to the ground state and taking absolute
squares results just in the exponential decay with rate $\bar{D}$.
This will be sufficient for the forthcoming considerations. For $t_f=2
\,k \,\pi, \,k 
=0,1,2,...$, both propagators are of course identical, due to the very
construction of the average.

When comparing these expressions with the numerical results described
later, we will see that they can account
for the monotonic background of the ionization rate (see figure
\ref{prominentsemi}), but
if we want to explain
the quasi--periodic modulations, responsible for the nonmonotonicity of
the rate, we have to go further in our semiclassical description.

\section{Semiclassical Propagators}

\subsection{General construction}
\label{prop}
To construct the semiclassical propagator outside the binding
potential, we
start with the (formal) path integral expression
\b U(x,t_f;y,t_i)&=&\int\limits_{t_i}^{t_f}
\mbox{$\cal D$}x(t)\,\exp\bigg({i\over h}S[x(t)]\bigg) \\
S[x(t)]&=&\int\limits_{t_i}^{t_f}L\bigg(x(t),\dot{x}(t)\bigg)\,dt\\
L&=&T-V=L_0-V_\delta=T-V_0-V_\delta \e
Here $T$ is the kinetic energy operator, $L$ is the full Lagrangian,
and $L_0$ is the Lagrangian for the electric field $V_0$ alone, with the
binding potential $V_\delta$ excluded.
The usual procedure in the semiclassical limit $h\to 0$ is to find the
stationary paths with $\delta S=0$, yielding the classical paths by
means of the Euler--Lagrange equation \cite{schulmanbook}.
The remarkable point here is that $V_\delta=-h\gamma\delta$ contains a
factor $h$. This becomes important for the semiclassical limit,
because this $h$ cancels in the exponent ${i S/ h}$, and consequently this
part of the phase does no more fluctuate arbitrarily fast for
non--stationary paths in the semiclassical limit.

Applying the analogy to saddle--point integration in function space
(cf. appendix \ref{spi}), we notice
that we only have to vary $S_0=\int\limits L_0$ in order to find the
stationary paths to $\delta S_0=0$. This condition gives the classical
path $x_{cl}(t)$ to $L_0$ by means of the Euler-Lagrange equation
\b {d\over dt}\bigg({\partial \over \partial \dot{x}}L_0
\bigg)-{\partial
\over \partial x}L_0=0 \ ,\label{ele0}\e
subject to the boundary conditions imposed by the path integral.
\b x_{cl}(t_i)&=&y\ \ ,\  x_{cl}(t_f)=x \\
x_{cl}(t)&=&x_{cl}(t|x,t_f;y,t_i) \\
 &=&-\cos(t) +\cos(t_i)+y+{x-y+\cos(t_f)-\cos(t_i) \over
 t_f-t_i}(t-t_i)\e
$V_\delta$ in the full Lagrangian $L$ only accounts for an
additional phase factor
$\exp(i\gamma\phi)$ to the propagator
\b \phi&=&\int\limits_{t_i}^{t_f}\delta\big(x_{cl}(t)\big)\,dt
\label{phase} \\
 &=&\sum\limits_{t_0^j}1/|\dot{x}_{cl}(t_0^j)|\ , \e
where the $t_0^j$ denote the zeros of the classical path
$x_{cl}(t_0^j)=0$. This phase $\phi$ jumps every time the classical
path $x_{cl}(t|x,t_f;y,t_i)$ crosses the $\delta$--potential at the
origin.

The result for the semiclassical propagator $U^{sc}$ is
\b U^{sc}(x,t_f;y,t_i)&=&{1 \over \sqrt{2\pi i
h}}\sqrt{-{\partial^2\over \partial x \, \partial y}S_0}\ \exp({i \over
h}S_0) \exp(i\gamma \phi)\\
&=&{1 \over \sqrt{2\pi i h (t_f-t_i)}}\exp \bigg( {i\over h}
\int\limits_{t_i}^{t_f}L_0(x_{cl}(t),\dot{x}_{cl}(t))\,dt \bigg)
\exp(i\gamma \phi) \label{usc} \e
The same result is derived in appendix \ref{wkb} using the
WKB--ansatz.

In general, one would have to include so--called Maslov phase factors
\cite{maslov}, but we can omit them because we do not encounter any
caustics in this problem. Since $V_0=-x\cos(t)$ is linear in $x$, the
appropriate semiclassical propagator for $L_0$ is identical
\cite{feynmanhibbs} to the exact one, namely the well--known Volkov
propagator $U^V$ \cite{volkov}.
The result can now be understood as the Volkov propagator $U^V$ plus
additional phase jumps for every crossing of the origin.
\b U^{sc}=U^V\,\exp(i\gamma \phi) \e

\subsection{Special tunneling propagator}
There is one important point to note; the description of the propagator
using regular classical paths is only justified after the electron has
tunneled out.
So in order to describe the tunneling paths, which classically do not
exist,
one has to modify the above description; a common method is to
introduce complex time and coordinates (cf. appendix \ref{complext} and
\cite{complextime,child}).

In our case, we know the bound state $\psi_0=\sqrt{\gamma /
h}\exp(-{\gamma}|x| /h)$, which formally equals a free wave $\exp({ipx
/ h})$ with
complex momentum $p_0=\pm i\gamma$. This is consistent with a negative
energy
$E=p_0^2/2=-\gamma^2/2$, which is just the ground state energy $E_0$ of
the $\delta$--potential.

We choose the following complex boundary conditions (a similar
reasoning appeared
in \cite{perelomovquasi})
for the complex tunneling path $x_T$:

the  initial momentum (imaginary part $\Im$ considered only)
\b \Im (\dot{x}_T(t_0)) =p_0=+i\gamma\e
(plus sign chosen to ensure exponential decay of wavefunctions
and not growth),\\
and the initial position
\b x_T(t_0)=y \label{second}\e
Additionally, $x_T$ must fulfill the final
condition \b x_T(t_f)=x\ ,\e
which is the boundary condition at the end of the path. The additional
free constant $t_0$ is necessary because we impose three boundary
conditions. But the ordinary differential equation
(\ref{ele0}), which $x_T$ must obey, is of order two,
and therefore has only two free constants.

The idea is that tunneling takes place in the imaginary part between
$t=t_0$
and $t=t_f$, and free propagation $U^{sc}$ (under the influence of the
oscillating electric field) takes place vice versa in the real part.
This interpretation is allowed by the usual decomposition rules for
semiclassical operators (cf. \cite{berrymount,voros}).

\begin{figure}[!htb]
\psfig{file=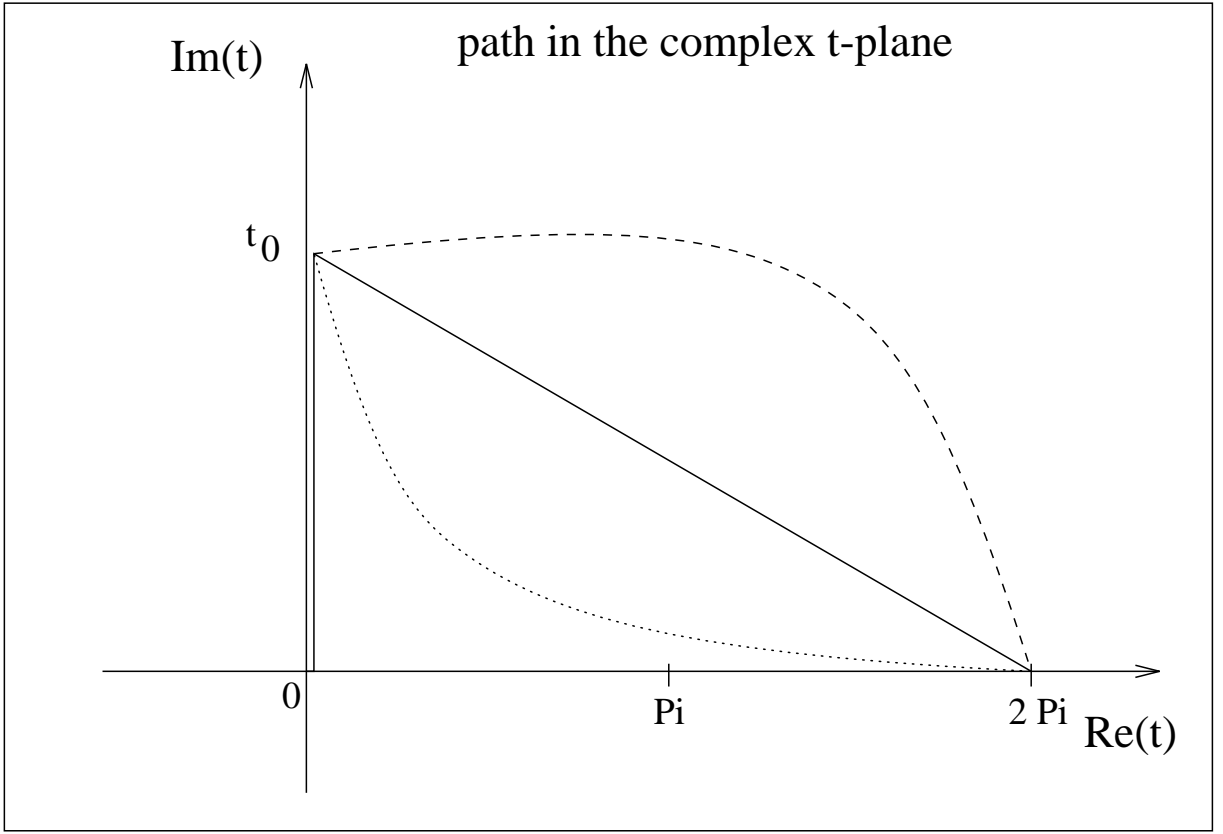,height=11cm,angle=270}
\caption[Path in the complex $  t$--plane, describing the
evolution of the first
wave packet stemming from $ { t=0}$. Since the line integral for
the calculation of the semiclassical propagator is path--independent,
these paths can be chosen at random.]
{Path in the complex $  t$--plane, describing the
evolution of the first
wave packet stemming from $ { t=0}$. Since the line integral for
the calculation of the semiclassical propagator is path--independent,
these paths can be chosen at random.
Important is that the propagation from $ {t=0}$ to
$ {t=t_0}$ uses
another propagator $ { \hat{\bar{ U}} ^\delta} $
than the propagator $ { \hat{ U}^T}$ afterwards to
$ {t_f=2\,\pi}$.}

\label{complexpath}
\end{figure}

The general expression for such a path $x_T$ fulfilling the (now
complex--valued) equation of motion (\ref{ele0}) is
\b x_T(t)&=& -\cos(t)+\cos(t_0)+y+v_0(t-t_0) \label{xtunnel}\\
   \dot{x}_T(t)&=& \sin(t)+v_0 \e
The first condition $\Im(\dot{x}_T(t_0))=p_0$ yields
\b i\gamma=\Im\bigg(\sin(t_0)+v_0\bigg) \e
Now we see the meaning of $t_0$; it must account for the complex
boundary condition and so we set
\b t_0=i\  \mbox{arcsinh}(\gamma)\ ,\label{t0} \e
in order to allow $v_0$ to remain real (see also \cite{perelomovquasi}
for this result). The second boundary condition (\ref{second}) is
fulfilled trivially by the ansatz (\ref{xtunnel}), and from the third
conditon we obtain
\b v_0={x-y+\cos(t_f)-\cos(t_0) \over t_f - t_0} \label{v0} \e
Using the equations (\ref{xtunnel},\ref{t0},\ref{v0}), we can easily
construct
the complete propagator $U^T$ containing tunneling as well as
propagation in the electric field. The result is just the analytic
continuation of our former
result $U^{sc}$ in eq. (\ref{usc})
\b U^T(x,t_f;y,t_0)={1 \over \sqrt{2\pi i h (t_f-t_0)}}\exp \Big( {i\over h}
\int\limits_{t_0}^{t_f}L_0(x_{T}(t),\dot{x}_{T}(t))\,dt \Big)
\exp(i\gamma \phi[x_T])\  \label{UT}\e
Note that the argument of the square root in the denominator is now
truely complex, so that we have an ambiguity in choosing a certain
sheat of the complex root. We decide to define
$\sqrt{r\exp(i\varphi)}=-\sqrt{r}\exp(i\varphi/2),\ \varphi \in [0,2\pi
]$.
Note further that the tunneling starts at
$t_0=i\ \mbox{arcsinh}(\gamma)$, and in order to reach $t_0$, we
first have to propagate the ground state from $t=0$ to $t_0$ using
$\hat{\bar{U}}^\delta(t_0)$, the analytic continuation in time of
(\ref{udeltabar}). This path in  the complex $t$-plane is depicted in
figure \ref{complexpath}.

According to the composition rule for propagators, the complete
propagator $\hat{U}^c$ for the ground state from $t=0$ to $t_f$ is
\b \hat{U}^c(t_f,0)=\hat{U}^T(t_f,t_0)\,\hat{\bar{U}}^\delta(t_0)\e
or in coordinate representation
\b \hat{U}^c(x,t_f;y,0)=\int\limits_{-\infty}^{+\infty}
\hat{U}^T(x,t_f;z,t_0)\,\hat{\bar{U}}^\delta(z,t_0;y,0)\,dz\e

\begin{figure}[t]
\psfig{file=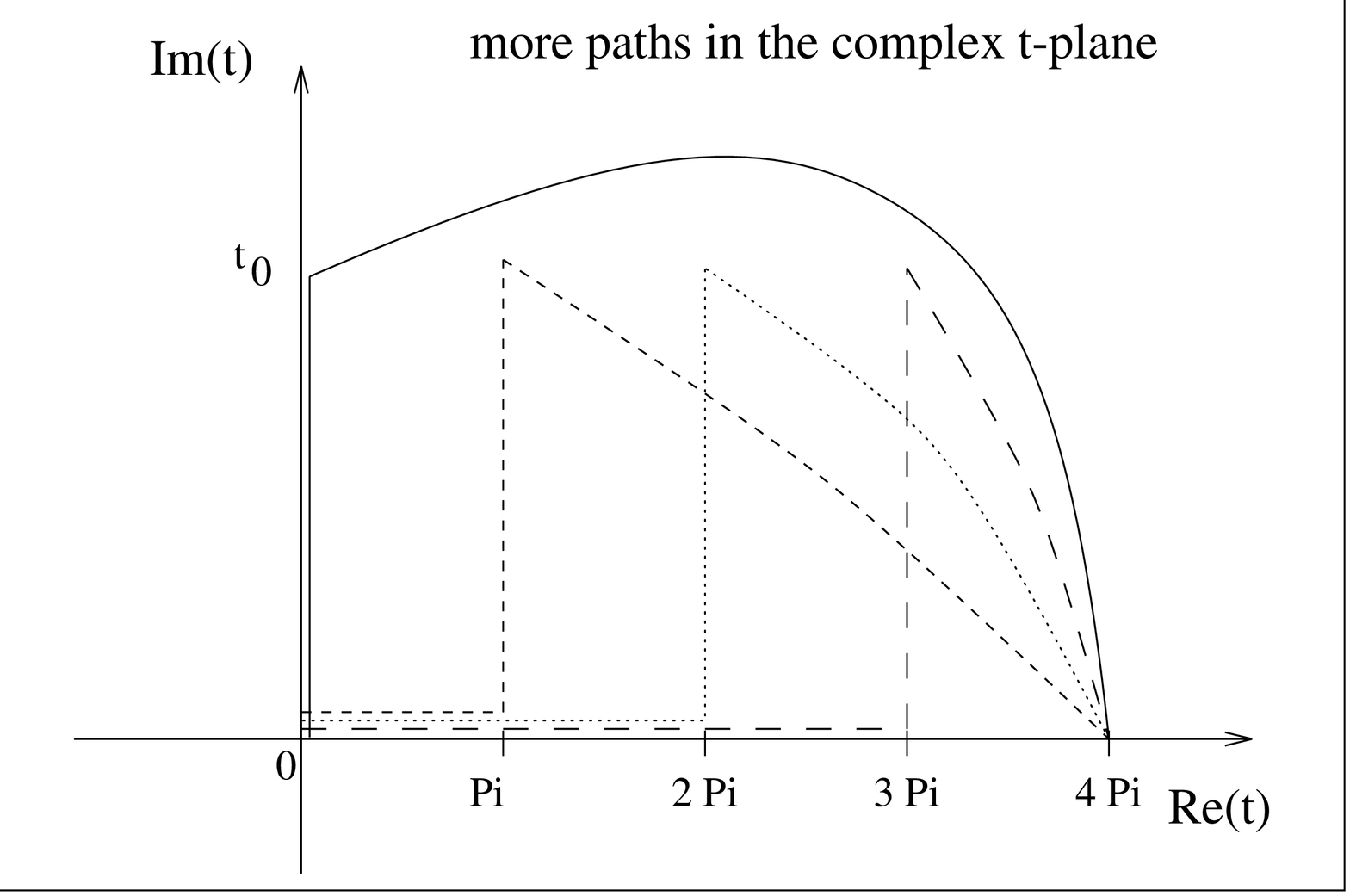,height=11cm,angle=270}
\caption{The various paths in the complex $  t$--plane for four
wave--packets are depicted.
The path for the wave packet emerging at $ {k\,\pi}$ goes from
$  0$ to
$ {k\,\pi +t_0}$ and afterwards to $ {t_f=4\,\pi}$. Again
we have path--independence for the two parts.}

\label{morepath}
\end{figure}

\subsection{Generalization to other ionization bursts}
The result of the previous section can be easily generalized to later
ionization bursts.
Figure \ref{complexpath} shows in the complex $t$--plane that
the electron does not become instantaneously free, but rather
propagates from
$t=0$ to $t=t_0$ according to the propagator
$\hat{\bar{U}}^\delta(t_0)$ (valid inside
the binding potential), and then tunnels and propagates from $t_0$ to
$t_f=2\pi$ according to the complex--valued propagator $\hat{U}^T$
valid outside. The above line integrals over the depicted paths are
path--independent \cite {complextime}, so there does not exist a unique
path.
The important point is the propagation with different propagators from
distinct starting points to distinct endpoints.
This describes the first ionization burst, but in order to describe the
wave packets emerging from the bursts at times $t_k=k\,\pi, k=0,
1,2,3,...,\ k\,\pi<t_f$,
one can repeat the above calculations.

The propagator $\hat{U}^c_k(t_f,0)$ for the wave packet stemming from
$t=k\pi$ is just
\b \hat{U}^c_k(t_f,0)=\hat{U}^T(t_f,k\pi+t_0)\,\hat{\bar{U}}^\delta
(k\pi+t_0)\e
 In this notation, $\hat{U}^c_0$ is identical to the above $\hat{U}^c$.
The interpretation is that the electron remains trapped from $t=0$ to
$t=k\pi$ and then propagates into the complex $t$--plane up to
$t=k\,\pi+t_0$, both according to $\hat{\bar{U}}^\delta$. It then
tunnels and propagates from $t=k\pi+t_0$
up to $t=t_f$, according to $\hat{U}^T$ (\ref{UT}), i.e. the (complexified) Volkov
propagator plus phase jumps.
Figure \ref{morepath} contains the four paths corresponding to four
wave packets created at $t=0,\pi,2\,\pi,3\,\pi$, which interfere at
$t_f=4\,\pi$.
Once again, these line integrals are path--independent so that one can
choose these paths at random.

\section{Interference between Paths}

The electron has two possibilities, it can tunnel and propagate as well
as remain bound by the binding potential. The quantum mechanical
amplitudes for both
processes are known and they can be added in order to obtain a better
description of the evolution of the system. This is just the ``semiclassical sum over classical paths'' method, e.g. \cite{gutzwillerbook}.

The full propagator $\hat{U}$ is the sum of the propagator
$\hat{\bar{U}} ^\delta$, valid for the part bound by the
$\delta-$potential, and the $\hat{U}^c_k$
's, the propagators for tunneling at $t=k\,\pi$ and (free) propagation
afterwards.
\b \hat{U}(t_f,0)=\hat{\bar{U}}^\delta(t_f) + \sum\limits_{k=0,1,2,...}
\hat{U}^c_k(t_f,0) \label{allesdrin}\e
\subsection{First period}
For simplicity and notational reasons, we will first consider only the
wave
packet originating from the ionization burst at $t=0$.
The other propagators can be added as well and the resulting integrals
can be evaluated using the same techniques as described below.
We will restrict ourselves to examining the wave function after full
periods
$t_f=2\,n\,\pi$, and we will deal mostly with just one period.
The second burst occuring at $t=\pi$ during the first period is of
secondary
importance, because this free electron follows a classical trajectory
$x_{cl}=a+b(t-\pi)-1-\cos(t)$ and its center is about $-2$ to the left
at $t_f=2\,n\,\pi$.
Therefore the overlap of this wave packet with the ground state
$\psi_0$ can be neglected.

This effect as well as the influence of considering several periods and
wave
packets will be demonstrated, when evaluating the analytic expressions
derived
below and comparing them to numerical results in section \ref{compnum}.

This propagator (\ref{allesdrin}) must be applied to the ground state
$\psi_0$ (\ref{psinull}) in order to obtain the
wavefunction $\psi(t_f)$ at a certain time $t_f$
\b \psi(t_f)=\hat{U}(t_f,0)\,\psi_0 \label{propagat}\e
Applying the propagator $\hat{\bar{U}}^\delta(t)$ results in just a
phase factor
$\exp(-(i\bar{E}^m /h)t)$, and so the above expression (\ref{propagat})
simplifies to (for one period $t_f=2\pi$)
\b \psi(x,t_f)=\exp(-{i\bar{E}^m \over h}t_f)\, \psi_0(x)    +
\exp(i\gamma\phi)\,\exp(-{i\bar{E}^m \over
h}t_0)\int\limits_{-\infty}^{+\infty}
U^V(x,t_f;y,t_0)\,\psi_0(y)\,dy\  \e
Using the property (\ref{spatialdelta}) that $\psi_0(y)$ approaches the
spatial $\delta$-function in the semiclassical limit $h\to 0$,
the integral can be evaluated to
\b \psi(x,t_f)&=&\exp(-{i\bar{E}^m \over h}t_f)\, \psi_0(x)    +
\exp(i\gamma\phi)\,\exp(-{i\bar{E}^m \over h}t_0)2\sqrt{h \over
\gamma}\,
U^V(x,t_f;0,t_0)\e

In order to calculate the probability amplitude $p$ for the electron to
remain bound, one has to project onto
the ground state $\psi_0(x)$
\b p= \int\limits_{-\infty}^{+\infty}\psi(x,t_f)\,\psi_0^*(x)\, dx\e
Using the spatial localization property
(\ref{spatialdelta}) again, this simplifies to
\b p=
\exp(-{i\bar{E}^m \over h}t_f) + \exp(-{i\bar{E}^m \over h}t_0)4{h
\over \gamma}
U^V(0,t_f;0,t_0) \label{oneperiod} \e
$\bar{E}^m$ and $t_0$ are known from the equations (\ref{barem}) and
(\ref{t0}),
respectively. The phase $\phi$ is identical to $0$ because the relevant
classical path $x_{cl}=1-\cos(t)$ never crosses the origin. The first
part of this expression for $p$ clearly accounts for the background,
while the second part determines the amplitude, phase, and period of the
superposed (slow) modulation.


\subsection{Fundamental periodicity}
The phase of the first term in eq. (\ref{oneperiod}) is mainly given
by the expression $-E_0\,t_f/h$, and the phase of the second term is dominated
by the action $S^{cl}/h$ along the classical path $x_{cl}=1-\cos(t)$.
\b
S^{cl}=\int\limits_0^{t_f}\bigg(\dot{x}_{cl}^2(t)+x_{cl}(t)\cos(t)\bigg) dt=
-z\ t_f  \e
The last identity is straight forward (also in
atomic units). Combining the phases 
and comparing to multiples of $2\,\pi$
results (for $t_f=2\,\pi$) exactly in equation (\ref{bilance})
with $E_{kin}=0$.
This is just the threshold condition for channel closing,  and implies
the same periodicity.
The period in $z$ is $\Delta z=1/(1+2\,\gamma^2)$,
if $\gamma$ is kept fixed,
or a period of $\Delta z=1$, if the depth of the binding potential $\alpha$
is kept fixed (see figure \ref{figpond} later).

\subsection{More periods}
Again this result
can be generalized easily to more bursts and longer final times
$t_f>2\,\pi$. Be $t_f=2\,n\,\pi$, then one has to sum over $2\,n$
bursts and amplitudes, and (this time written explicitly) the result for
$p$ is
\b p&=&\exp(-{i\over h}\bar{E}^m t_f )+\sum\limits_{k=0}^{2\,n-1}\
{- 4 h \over g\,\sqrt{2 i\pi h ( t_f -t_0-k\pi)}}
\exp\bigg(  \zeta^k\bigg) \label{result}\\
\zeta^k&=&{-i \over 4 h( t_f -t_0-k\pi )}\bigg( (t_0+k\pi)^2 +
(t_f-t_0-k\pi)\cos(t_0)\sin(t_0) \\
\nonumber && -2(t_0+k\pi) t_f
 +4\cos(t_0)(-1)^k-2+ t_f ^2
-2\cos^2(t_0)\bigg) \
 -{i\over h}\bar{E}^m(t_0+k\pi)       \e
Note the useful relations $\cos(t_0)=\sqrt{1+\gamma^2}$ and
$\sin(t_0)=i\,\gamma$. Propagating the system for longer times $t_f>2\,\pi$
means to have more phase built up in the exponents of (\ref{result}),
and results in a finer resolution in the ionisation rate $\Gamma$; higher
frequencies than the basic modulation period can be accounted for.
How this can explain the fine--structure is demonstrated later on in
figure \ref{numericsfine}.

\clearpage

In the above sum (\ref{result}), only the contributions with
$k$ even are important.
This is because the centers of wave packets stemming from $k$ odd are
located at about $-2$ to the left at
$t_f=2\,n\,\pi$, and therefore the overlap is very small.

\subsection{Ionization rate}
The probability $w$ for not being ionized
is now calculated by taking the absolute square $w=|p|^2$, and the
corresponding
ionization rate $\Gamma$, fitting the exponential decay $w=\exp(-\Gamma\,
t_f/(2\pi))$, can be defined as
\b \Gamma=-{2\pi \over t_f}\ln (|p|^2) \label{gamma} \e
For this quantity $\Gamma$, we will now compare the above semiclassical
results with those from numerical simulations \cite{ksdiss}.

\clearpage
\section{Comparisons}
\subsection{Comparison with numerical results}
\label{compnum}

\begin{figure}[htb]
\psfig{file=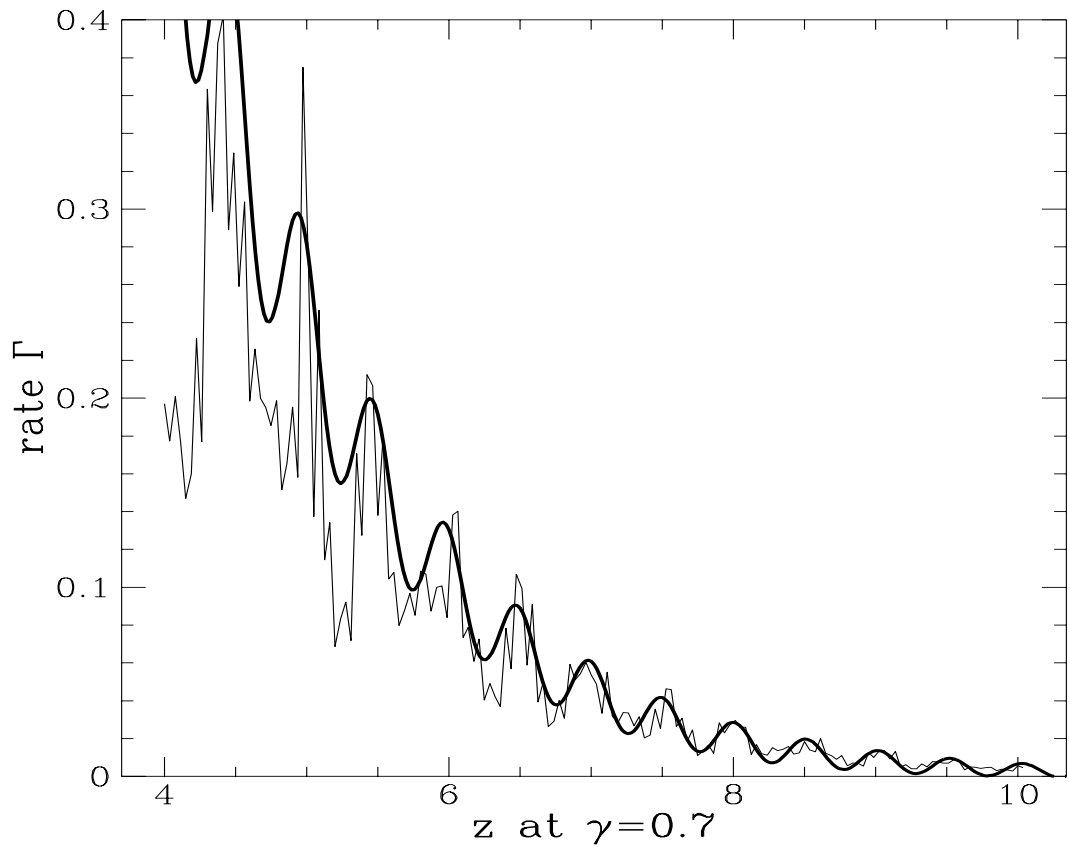,width=15cm,height= 12cm}
\caption{Numerical ionization rate (thin and jagged) and semiclassical
approximation (thick and smooth) versus number of
ponderomotive photons $ {} z$ for $ {}{ \gamma=0.7}$}

\label{numerics1}
\end{figure}

\begin{figure}[!tb]
\psfig{file=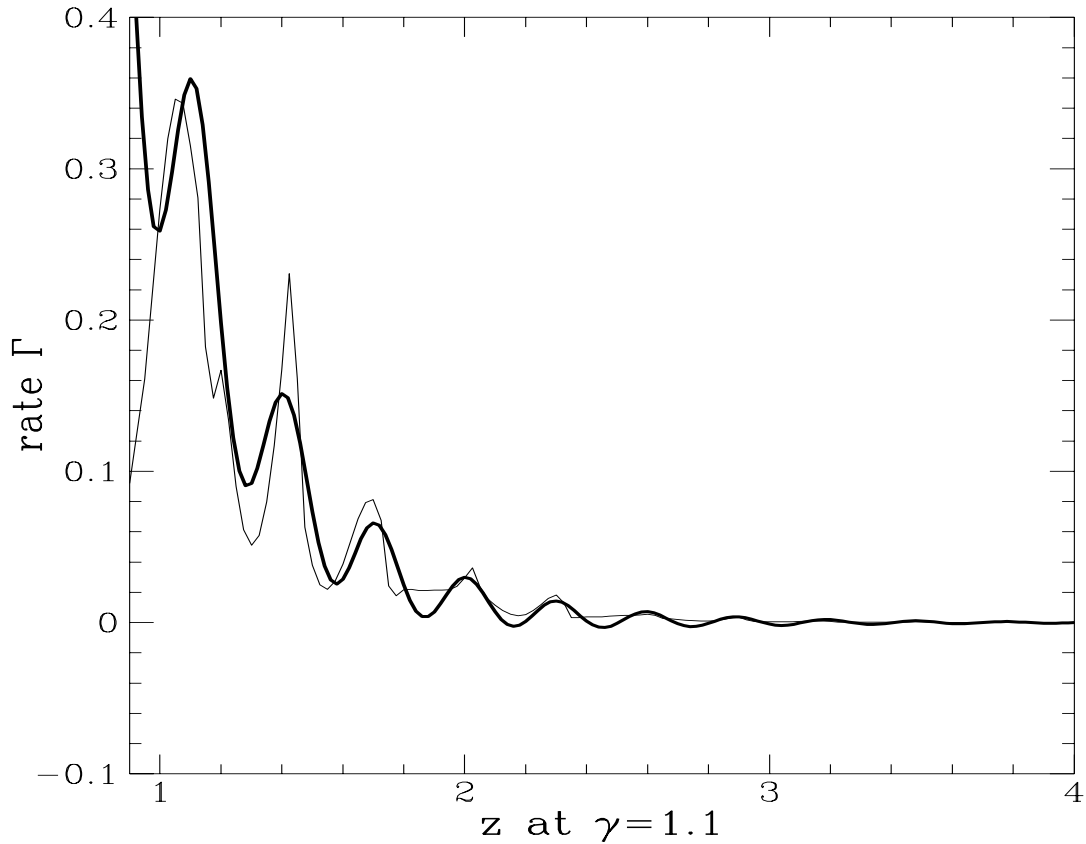,width=15cm,height= 12cm}
\caption{Numerical ionization rate (thin and jagged) and semiclassical
approximation (thick and smooth) versus number of
ponderomotive photons $ {} z$ for $ {}{\gamma=1.1}$}
\label{numerics2}
\end{figure}

The numerical results are obtained by an integral equation method
implemented by
K.\, Sonnenmoser (\cite{ksdiss} for details), which allows
high--resolution scans and exhibits a lot of fine structure.
While $\gamma$ is kept fixed, $z$ is varied, and so the semiclassical
limit $h \to 0$ corresponds just to $z \to \infty$.
This means that the agreement will become the better the larger $z$
is.

The interesting region for $\gamma$ is of course $\gamma\approx 1$,
because
this is the transition region between adiabaticity and multiphoton
regime.
For $\gamma\ll 1$, the adiabaticity criterion is fulfilled, the
averaged WKB--value can be justified, and is in good agreement due to the
very construction of our theory.
For $\gamma \gg 1$, one should turn over to a pure multiphoton
description \cite{delonekrainov,faisalbook}.

Figure \ref{numerics1} shows the numerical result (thin and jagged) for
$\gamma=0.7$,
together with the results of our theory.
The background as well as the amplitude, phase, and periodicity of the
modulation are very well comprised in the
semiclassical theory for $z$ not too small.

The same is done in figure \ref{numerics2} for $\gamma=1.1$ . Here
again, one recognizes
 that the characteristic elements of the ionization curves
are in good agreement. The same is valid  

\begin{figure}[tb]
\psfig{file=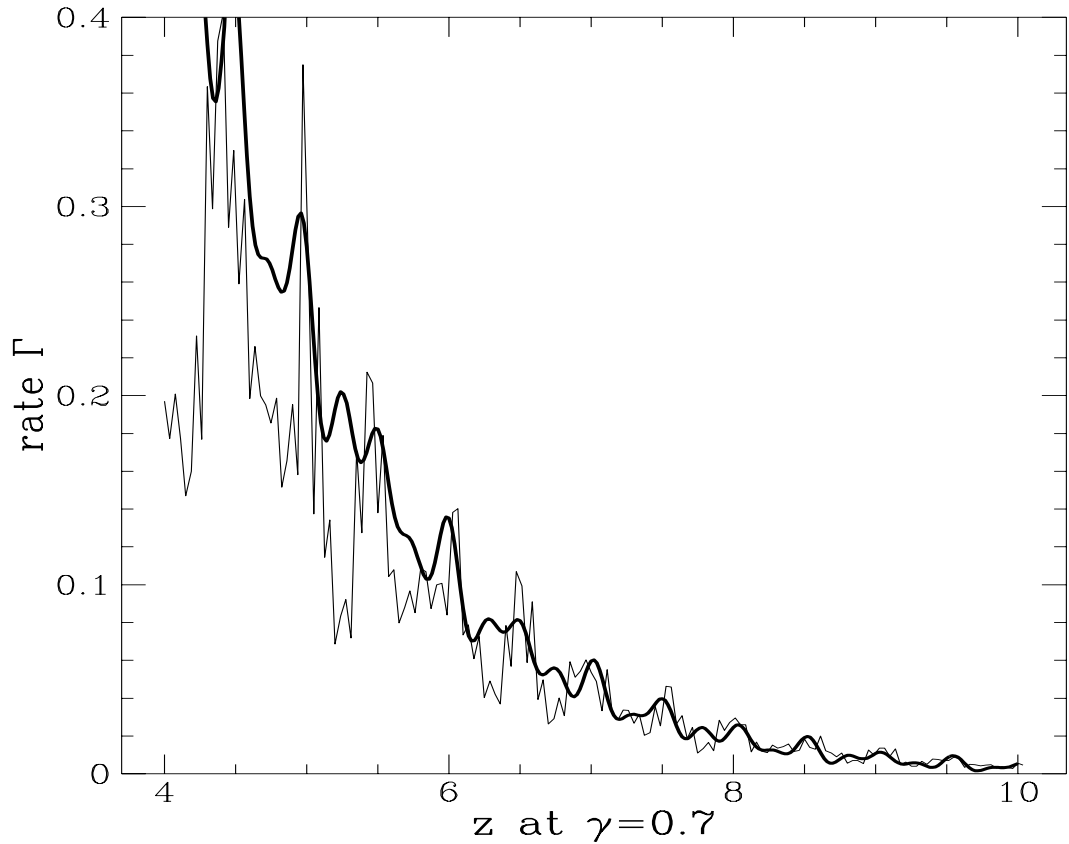,width=15cm,height= 12cm}
\caption{More fine--structure by considering longer periods than
$ {}{2\,\pi}$, here two cycles $ {}{4\,\pi}$ considered}
\label{numericsfine}
\end{figure}

\begin{figure}[tb]
\psfig{file=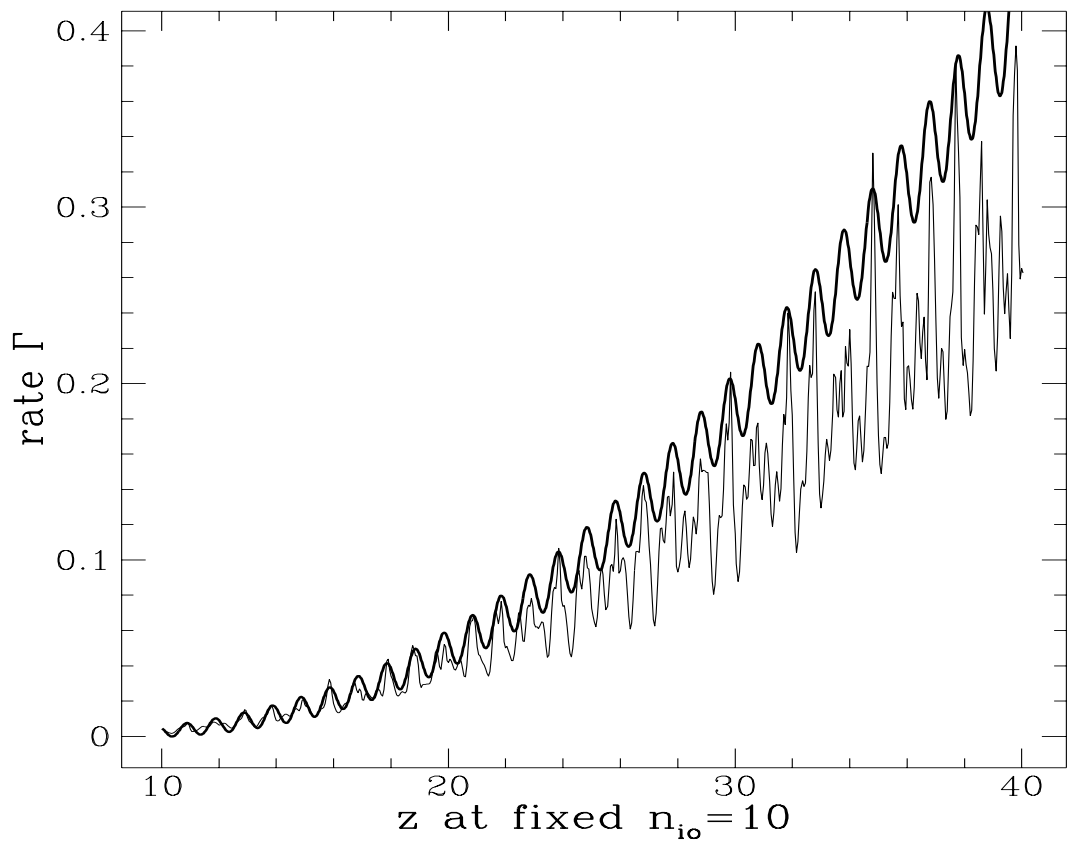,width=15cm,height= 12cm}
\caption{Now the depth of the binding potential is kept fixed, and
$ {} z$ is varied.
The semiclassical theory (thick smooth curve) is compared to numerical
results (thin and jagged).}
\label{figpond}
\end{figure}

of course for all other values
$\gamma\approx 1$ and can be extended up to $\gamma\approx 2.5$,
clearly beyond the adiabaticity regime.

If we want to incorporate more fine--structure superposed onto the
modulation, we can consider longer periods than $2 \pi$. The result is
a behaviour as in
figure \ref{numericsfine}, which resembles closely the fine--structure,
even if
there can be no one--to--one correspondence between every small wiggle.
Here
the wave packets stemming from ionization bursts $\pi$ and $3\,\pi$
were also
taken into account, but they have very little influence on the result
for $t_f=
4\,\pi$.
The main contributions come of course from the wave packets stemming
from $t=0$ and $t=2\,\pi$.
This can be seen
as an example for the above qualititative statement about the relative
importance of ionization at even or odd multiples of $\pi$.

The result (\ref{result}) for fixed $\gamma$ and $z$ can easily be
transcribed to other parameter combinations. A common representation of
ionization rates $\Gamma$ is to keep $n_{io}=2\gamma^2
z=\alpha^2/(2\omega)$ fixed, i.e. the depth of the binding potential
(last expression in atomic units as in section \ref{model}).
One varies the intensity $\mu^2$ at fixed frequency $\omega$, which
corresponds better to experimental situations. If one plots $\Gamma$
over the intensity or over $z=\mu^2/(4\omega^2)$, one again obtains
periodicity in the ponderomotive channel closing \cite{eberlychannel}.
This is depicted in figure \ref{figpond} in comparison with
numerical data; note hereby that the semiclassical limit
no more corresponds simply to $z\to\infty$, but is more involved and
cannot be included simply into the representation. One again notices
that the characteristic elements of ionization curves can be calculated in
the semiclassical theory.

\subsection{Comparison to other theories}

One major advantage and distinction of our theory from others (even
from those claiming to be semi-- or quasiclassic in some sense) is
that we encounter
no divergencies at channel closing thresholds, a result strongly
supported by numerical evidence. Such divergencies typically occur in
theories
that (in our mind somehow arteficially) separate the ionization process
into different channels, each one related to ionization by a distinct
number of photons. Every time a channel closes, the corresponding
ionization rate of the next higher channel (and so the overall rate)
becomes infinite.

\begin{figure}[tb]
\psfig{file=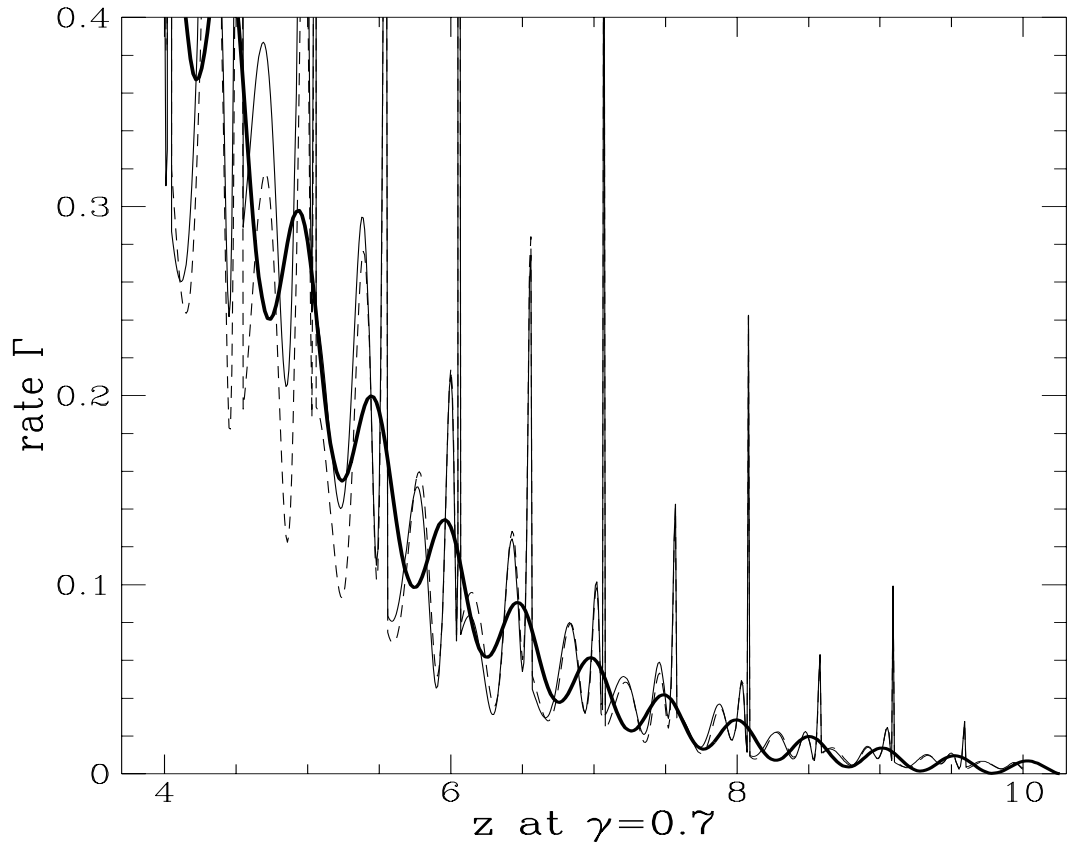,width=15cm,height= 12cm}
\caption[Comparison with other theories containing divergencies at
channel closing thresholds]{Comparison with other theories containing divergencies at
channel closing thresholds: semiclassical theory (thick and smooth)
versus Perelomov, Popov, Terent'ev [13]
(thin line) and Susskind, Cowley, Valeo [14]
(dashed), which nearly coincide in this region and especially in their
behaviour at thresholds.}
\label{divergencies}
\end{figure}

One earlier representative (eq.[31] by Perelomov, Popov,
Terent'ev \cite{perelomovbasic}; including quasiclassical features)
and one more recent representative
(eq.[44] by Susskind, Cowley, Valeo \cite{susskind};
asymptotic in the number of photons for ionization) 
of this kind of theory is shown in figure \ref{divergencies}, in
comparison with our result. The spikes in this figure are related
to the closing of certain ionization channels with, say, $k$ photons
at $z=z_k$ (eq. (\ref{zks})), and they can be traced back to the
divergence in the next higher channel with $k+1$ photons. 

In between these thresholds, the background and the imposed modulation
is approximately the same. But in contrast to these more implicit
theories, where results are quite involved, our result
(\ref{result},\ref{gamma}) allows the direct, separate, and explicit
evaluation of the background rate as well as of
the amplitude and phase of the modulation.

\clearpage

\section{Summary}
We succeeded in a semiclassical description of time--dependent
tunneling and ionization in an oscillating field.
The characteristic features of typical ionization
curves can now be explained using a picture of tunneling, propagating and
interfering wave packets (formula (\ref{result})).
The main ingredients are first the separation of the ionization process
into two disctinct steps, motivated by the asymptotic evaluation of
instantaneous WKB--rates.
And second the usage of complex time and analytical continuation of
propagators, necessary to account for tunneling by a classical path
description.
The slow modulation with channel closing periodicity (an idea stemming
from a
multiphoton viewpoint) can be described correctly with respect to
amplitude and to phase. Even the rich superposed fine--structure can be
accounted for by considering a multitude of interfering wave packets.

\section{Acknowledgments}
The numerical calculations used for comparison were done with a program
packet
\cite{ksdiss} kindly supplied by K. Sonnenmoser, whom is also thanked
for
two valuable discussions.

G. Scharf is thanked for valuable discussions and suggestions
at all stages of this work.

This work was supported by Schweizerischer Nationalfonds.

\clearpage
\begin{appendix}

\section{Saddle--Point Integration Techniques}
\label{spi}
Under the notation of saddle--point integration, there exist two
different versions in the literature (all considered in the limit $h
\to 0$):

{\bf Method of steepest descent:}
\b \int\limits g(t)\,\exp(-{1\over h }f(t))\, dt \approx\sqrt{2\pi h 
\over {\partial
^2 \over \partial ^2 t}f(t_0)}g(t_0)\,\exp( -{1\over h}f(t_0))
\label{steep}\\
 {\partial \over \partial t}f(t_0)=0 \ ,\      {\partial ^2 \over
 \partial t^2}f(t_0)>0 \e
Here the argument in the exponent was expanded around the minimum $t_0$
of $f(t)$, up
to second order, and the so created gaussian integral was evaluated
exactly. If there exist several minima, one has to take the sum over
them.

{\bf Method of stationary phase:}
\b \int\limits g(t)\,\exp({i\over h }f(t))\, dt \approx\sqrt{2\pi i h \over
{\partial ^2 \over \partial t^2}f(t_0)}g(t_0)\,\exp( {i\over h}f(t_0))\e
This is in some respect the analytic continuation of (\ref{steep});
here every
stationary point $t_0$ with $\partial_t\,f(t_0)=0$ is relevant, not
only the minima.

{\bf Analogy in functional integration:}
\b \int\limits_{t_i}^{t_f} \mbox{$\cal D$}x(t)\,g(x(t))\,\exp({i\over
h}S[x(t)]) \e
In this path integral, one has to integrate over all possible paths
with the
boundary conditions
\b x(t_i)=y\ ,\ x(t_f)=x_f \e
In the semiclassical approximation, one looks for the paths $x_{cl}$
which make the functional $S[x]$ stationary:
\b \delta S[x(t)]\Big|_{x_{cl}}=0 \e
One again expands around these classical paths $x_{cl}$ (cf. e.g.
\cite{schulmanbook}) and obtains the so--called
Van--Vleck propagator \cite{vanvleck}, which in general must be
corrected by
so--called Maslov factors $\nu$ \cite{maslov}. These take into account the
crossing of caustics, i.e. divergencies of the mixed second derivative of
$S^{cl}$.
\b {1\over \sqrt{2 \pi i h}}\sqrt{-{\partial ^2 \over \partial x_f \,
\partial y}S^{cl}}\,\exp({i\over
h}S^{cl})\int\limits_{t_i}^{t_f}g(x_{cl}(t))\,dt\ \exp(-i\,{\pi\over
2}\,\nu) \\
x_{cl}(t_i)=y\ ,\ x_{cl}(t_f)=x_f \ ,\ \delta S[x(t)]\Big|_{x_{cl}}=0
\ ,\ S^{cl}=S[x^{cl}]\e

\section{Semiclassical Propagator using the WKB--Ansatz}
\label{wkb}
This is an alternative to the construction of the semiclassical
propagator $U^{sc}(x,t;y,t_i)$ by the path intgral approach (cf.
section \ref{prop} and appendix \ref{spi}).
We construct $U^{sc}(x,t;y,t_i)$ in such a way that it
fulfills the time--dependent Schr\"odinger equation (\ref{transformed})
with respect to $t$ and $x$ up to $O(h^2)$.
\b i h {\partial \over \partial t}U^{sc}=\hat{H} U^{sc}&=&\bigg(-{1
\over 2}h^2  {\partial ^2 \over \partial x^2}-h \gamma \delta (x)- x
\cos(t)\bigg)U^{sc} \\
\lim_{t\to t_i} U^{sc}(x,t;y,t_i)&=&\delta (x-y) \label{unorm}\e

We make the ansatz
\b U=\varphi(x,t;y,t_i)\exp\bigg({i\over h}S(x,t;y,t_i)\bigg) \e
and yield
\b ih{\partial_t}\varphi -\varphi{\partial_t}S={1\over
2}({\partial_x}S)^2-ih{\partial_x}\varphi {\partial_x}S -{ih\over
2}\varphi{\partial_x^2}S -{h^2\over 2}{\partial_x^2}\varphi
-h\gamma\delta(x)\varphi-x\cos(t)\varphi \e
Comparing the distinct orders of $h$:
\b h^0&:& 0=\partial_t S+{1\over 2}(\partial_x
S)^2-x\cos(t)\label{h0}\\
   h^1&:& i\gamma\delta(x)\varphi=\partial_t\varphi + \partial_x\varphi
   \partial_x S + {1\over 2}\varphi\partial_x^2S \label{h1}\\
& & 2i\gamma\delta(x)\varphi^2={\partial \over \partial
t}\varphi^2+{\partial \over \partial x}(\varphi^2\partial_xS)
\label{consflow}\e
The right--hand side of equation (\ref{consflow}) is just a
conservation equation with density $\varphi^2$ and flow
$\varphi^2\partial_xS$, which is fulfilled everywhere except the
origin.
The equation (\ref{h0}) for $h^0$ is easily solvable, it is just the
Hamilton--Jacobi
equation $-\partial_t S=H(x,\partial_x S)$ for the motion of an
electron in an oscillating electric field (with Lagrangian $L_0$).
One can easily solve the
appropriate equation of motion and yields the classical path
$x_{cl}(\tau)$ with boundary conditions
$x_{cl}(t)=x$ and $x_{cl}(t_i)=y$.
\b x_{cl}(\tau)=-\cos(\tau) +\cos(t_i)+y+{x-y+\cos(t)-\cos(t_i)
\over t-t_i}(\tau-t_i)\e
The action $S$ is then the time
integral over the classical Lagrangian $L_0$:
\b S=\int\limits_{t_i}^tL_0\bigg(x_{cl}(t'),\dot{x}_{cl}(t')\bigg)\,
dt' \e

Then, as usual, the expression
$\varphi_0=\sqrt{-\partial_x\partial_yS}=1/\sqrt{t-t_i}$ solves the
equation (\ref{h1}) for $h^1$ outside the origin. Now we make the
ansatz
\b \varphi=\beta(x,t;y,t_i)\,\varphi_0={\beta\over \sqrt{t-t_i}}\e
Inserting this in (\ref{h1}), we yield
\b i\gamma\delta(x)\beta=\partial_t\beta + \partial_x\beta\,\,\partial_x
S\e
This linear partial differential equation of first order can be solved
using the method of characteristics. The (ordinary) differential
equation for the characteristic $x_T$ is
\b {d\over dt}x_T=\partial_xS(x,t;y,t_i) \e
But $S$ is, as we know, the action for the classical path and therefore
$\partial_xS$ is just the momentum of the classical path
$\partial_xS=\dot{x}_{cl}(t)$ (e.g. \cite{goldstein}).
So we conclude that the characteristic is just the classical path
$x_T=x_{cl}$.
We obtain the following (ordinary) differential equation for $\beta$ :
\b {d\over
dt}\beta\Big|_{x_{cl}}=\partial_t\beta+\partial_x\beta\,{d\over
dt}x_{cl}=i\gamma\delta(x)\,\beta\Big|_{x_{cl}}\e
This can be integrated straight forward and the result is
\b \beta(t)\Big|_{x_{cl}}=\exp\Big(i\gamma
\int\limits_{t_i}^t\delta(x_{cl}(t'))\,dt'\Big)\,\beta_0 \e
This again can be understood as a phase jump for every crossing of the
$\delta$--potential at the origin; the constant $\beta_0$ is fixed to
$\beta_0=1/\sqrt{2\pi i h}$ by the required normalization (\ref{unorm})
of $U^{sc}$.

To summarize, we end up with the following result, identical to eq. (\ref{usc})
\b U^{sc}(x,t;y,t_i) ={1 \over \sqrt{2\pi i h (t-t_i)}}\exp \bigg(
{i\over h} S[x_{cl}](x,t;y,t_i) \bigg)  \exp\bigg(i\gamma
\int\limits_{t_i}^t\delta(x_{cl}(t'))\,dt'\bigg) \e

\section{Complex Time Description of Tunneling Processes}
\label{complext}
The usefulness of a complex time coordinate for describing tunneling
processes
is best demonstrated by a simple example. Consider an inverse harmonic
potential
barrier $V(x)=-x^2/2$ and a particle coming in from $-\infty$ with
energy $E=-1/2$. The classical equation of motion
$\stackrel{..}{x}_{cl}=x_{cl}$ is fulfilled by the classical trajectory
$x_{cl}(t)=-\cosh(t)$, restricting
the particle to $x_{cl}\leq -1$.

\begin{figure}[!tb]
\psfig{file=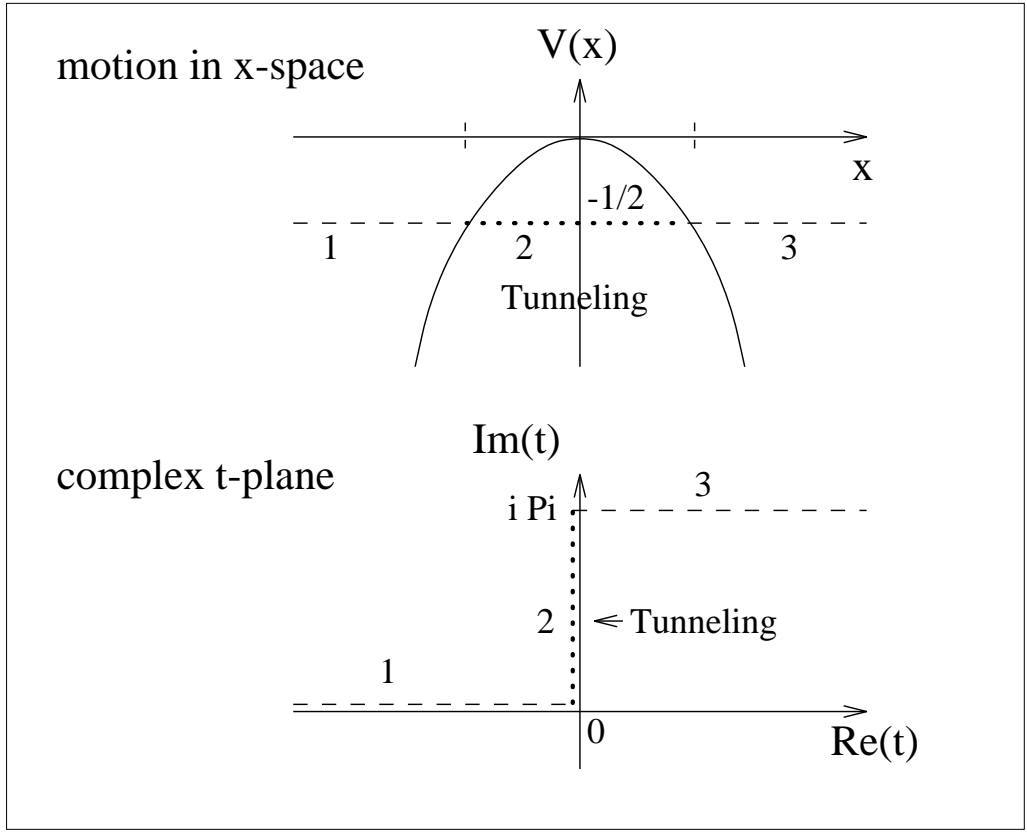,height=11cm,angle=270}
\caption{Tunneling in $  x$--space and the appropriate complex
time path for an inverse harmonic potential}

\label{barrier}
\end{figure}

In this time--independent problem, the total energy $E$ is a constant
of motion:
\b E={\dot{x}^2\over 2}-{1\over 2}x^2 = - {1 \over 2}\e
This differential equation can be solved for $t(x)$:
\b t(x)=\pm\int\limits_{-1}^x{dx'\over \sqrt{2E+x'^2}} \e
The range of $x$ can now be extended formally to $x>-1$, then $t$
acquires an imaginary part for $x'\in [-1,1]$
\b \int\limits_{-1}^1{dx'\over \sqrt{2E+x'^2}}=i\pi \e
The tunneling trajectory through the barrier can be
described in the complex $t$--plane
by the path $[-\infty , 0],[0 , i\,\pi],[i\,\pi , +\infty +i\,\pi]$.
The second part
is the tunneling process through the barrier, noting that
$$x_{cl}(i\tau)=-\cosh(i\tau)=
-\cos(\tau)$$
The third part describes the propagation on the other
side of the barrier, noting that
$$x_{cl}(t+i\pi)=-\cosh(t+i\pi)=+\cosh(t)$$
\clearpage
Figure  \ref{barrier}
illustrates this by contrasting the path in $x$--space to the path
in the complex $t$--plane.
For mathematical foundations of the analytic continuation applied above
see \cite{complextime}, for recent applications of this method compare
e.g. \cite{child,finnland}.

\clearpage
\section{Extensibility and Comparison to other Models}
In section \ref{trafo}, there was explicitly made use of the scaling property
$\delta(\beta\, x)
=\delta(x)/\beta$ (for $\beta >0$) of the one--dimensional
$\delta$--function.
For the Coulomb--potential $V_C$, the scaling behaviour is identical, so that
the whole calculation can be repeated. The main difference consists
in a different phase factor $\phi_C$ (cf. (\ref{phase})), which is now
the line integral over the Coulomb--potential:
\b \phi_C&=&\int\limits_{t_i}^{t_f}V_C\big(x_{cl}(t)\big)\,dt \e
This is due to the fact that the binding potential is suppressed by a
factor $h$
in the transformed Schr\"odinger equation (\ref{transformed}). As
derived in  sections \ref{prop} and appendix \ref{wkb}, such a 
suitably suppressed binding potential
does not influence the classical trajectories, but only changes the
phase transported along these trajectories by an additional phase
factor $\phi_C$ (given above).

This separation is no longer valid for other types of potentials with
different (or without) scaling properties. Here the classical
equations of motion must be solved fully for the binding potential plus
electromagnetic field.
In the time--dependent case, there generally does not exist a first
integral of motion (like the energy in the static case), and so the
classical system is not integrable in closed form.

The physics of the ionization process should remain the same: complex
time--dependent tunneling followed by free propagation (two--step
models). This is reflected by the fact that
ionization curves for other model potentials show qualitatively the
same characteristics.
The semiclassical ``sum over classical paths''--method claims to work
in this case anyway, but unfortunately this cannot be done in
analytical form because of the nonintegrability already on the classical
level.

\end{appendix}


\clearpage

\begin{thebibliography}{10}

\bibitem{ksdiss}
{G. Scharf, K. Sonnenmoser, W.F. Wreszinski}, {Phys. Rev. A} {\bf 44},  3250
  (1991).

\bibitem{mainfray}
{G. Mainfray, C. Manus}, Rep. Prog. Phys. {\bf 54},  1333  (1991).

\bibitem{knightreview}
{K. Burnett, V.C. Reed, P.L. Knight}, J. Phys. B: At. Mol. Opt. Phys. {\bf 26},
   561  (1993).

\bibitem{eberlyrev}
{J.H. Eberly, J. Javanainen, K. Rzazewski}, Phys. Rep. {\bf 204},  331  (1991).

\bibitem{gavrila}
{\em Atoms in Intense Laser Fields}, edited by M. Gavrila (Academic, Orlando,
  1992).

\bibitem{special}
special issue of J. Opt. Soc. Am. B {\bf 7},    ({4, 1990}).

\bibitem{ks2}
K. Sonnenmoser, J. Phys. B: At. Mol. Opt. Phys. {\bf 26},  457  (1993).

\bibitem{negions}
H. Massey, {\em Negative Ions} (Cambridge University Press, Cambridge, 1976).

\bibitem{keldysh}
L. Keldysh, Sov. Phys. JETP {\bf 20},  1307  (1965).

\bibitem{faisal}
F. Faisal, J. Phys. B {\bf 6},  L89  (1973).

\bibitem{zreiss}
H. Reiss, Phys. Rev. A {\bf 22},  1786  (1980).

\bibitem{volkov}
D. Volkov, Z. Phys. {\bf 94},  250  (1935).

\bibitem{perelomovbasic}
{A.M. Perelomov, V.S. Popov, M.V. Terent'ev}, Sov. Phys. JETP {\bf 23},  924
  (1966).

\bibitem{susskind}
{S.M. Susskind, S.C. Cowley, E.J. Valeo}, Phys. Rev. A {\bf 42},  3090  (1990).

\bibitem{scully}
{W. Becker, R.R. Schlicher, M.O. Scully}, J. Phys. B {\bf 19},  L785  (1986).

\bibitem{kulander}
{K.C. Kulander, K.J. Schafer, J.L. Krause},  in {\em Proceedings of the
  Workshop Super Intense Laser Atom Physics SILAP III}, {\em NATO Advanced
  Study Institute Series B}, edited by {B. Piraux et al} (Plenum Press, New
  York, 1993).

\bibitem{corkum}
P. Corkum, Phys. Rev. Lett. {\bf 71},  1994  (1993).

\bibitem{lewenstein}
{M. Lewenstein, Ph. Balcou, M.Y. Ivanov, A. L'Huillier, P.B. Corkum}, Phys.
  Rev. A {\bf 49},  2117  (1994).

\bibitem{shirley}
J. Shirley, Phys. Rev. B {\bf 138},  979  (1965).

\bibitem{chu}
S. Chu, Adv. At. Mol. Phys. {\bf 21},  197  (1985).

\bibitem{floquetposha}
{R.M. Potvliege, R. Shakeshaft}, Phys. Rev. A {\bf 38},  4597  (1988).

\bibitem{beckerdreid}
{W. Becker, J.K. McIver, M. Confer}, Phys. Rev. A {\bf 40},  6904  (1989).

\bibitem{delonekrainov}
{N.B. Delone, V.P. Krainov}, {\em Multiphoton processes in atoms} (Springer,
  New York, 1994).

\bibitem{faisalbook}
F. Faisal, {\em Theory of Multiphoton Processes} (Plenum Press, New York,
  1987).

\bibitem{recipes}
{W.H. Press, S.A. Teukolsky, W.T. Vetterling, B.P. Flannery}, {\em {Numerical
  Recipes, second edition}} (Cambridge University Press, Cambridge, 1992).

\bibitem{eberlychannel}
J. Eberly, J. Phys. B: At. Mol. Opt. Phys. {\bf 23},  L619  (1990).

\bibitem{ebergreen}
{W.G. Greenwood, J.H. Eberly}, {Phys. Rev. A} {\bf 43},  525  (1991).

\bibitem{gutzwillerbook}
M. Gutzwiller, {\em Chaos in Classical and Quantum Mechanics} ({Springer}, New
  York, 1990).

\bibitem{elberfeldkleber}
{W. Elberfeld, M. Kleber}, Z. Phys. B {\bf 73},  23  (1988).

\bibitem{pontshake}
{M. Pont, R. Shakeshaft, R.M. Potvliege}, Phys. Rev. A {\bf 42},  6969  (1990).

\bibitem{geltman}
S. Geltman, J. Phys. B. {\bf 11},  3323  (1978).

\bibitem{schulmanbook}
L. Schulman, {\em Techniques and Applications of Path Integration} (John Wiley
  \& Sons, USA, 1981).

\bibitem{maslov}
{V.P. Maslov, M.V. Fedoriuk}, {\em Semi--classical Approximation in Quantum
  Mechanics} ({ D. Reidel Publishing Company}, Dordrecht, 1981).

\bibitem{feynmanhibbs}
{R.P. Feynman, A.R. Hibbs}, {\em Quantum mechanics and Path Integrals}
  (McGraw--Hill, USA, 1965).

\bibitem{complextime}
D. McLaughlin, J. Math. Phys. {\bf 13},  1099  (1972).

\bibitem{child}
M. Child, {\em Semiclassical Mechanics with Molecular Approximations}
  (Clarendon Press, Oxford, 1991).

\bibitem{perelomovquasi}
{V.S. Popov, V.P. Kuznetov, A.M. Perelomov}, Sov. Phys. JETP {\bf 26},  222
  (1968).

\bibitem{berrymount}
{M.V. Berry, K.E. Mount}, Rep. Prog. Phys. {\bf 35},  315  (1972).

\bibitem{voros}
A. Voros, Ann. Inst. Henri Poincar\'e {\bf XXIV},  31  (1976).

\bibitem{vanvleck}
J. van Vleck, Proc. Natl. Acad. Sci. USA {\bf 14},  178  (1928).

\bibitem{goldstein}
H. Goldstein, {\em Classical Mechanics} (Addison--Wesley, Cambridge, Mass.,
  1950).

\bibitem{finnland}
{M. Kira, I. Tittonen, W.K. Laii, S. Stenholm}, Phys. Rev. A {\bf 51},  2826
  (1995).

\end{thebibliography}
\end{document}